\documentclass{article}
\usepackage{PRIMEarxiv}
\usepackage[utf8]{inputenc} % allow utf-8 input
\usepackage[T1]{fontenc}    % use 8-bit T1 fonts
\usepackage{hyperref}       % hyperlinks
\usepackage{url}            % simple URL typesetting
\usepackage{booktabs}       % professional-quality tables
\usepackage{amsfonts}       % blackboard math symbols
\usepackage{nicefrac}       % compact symbols for 1/2, etc.
\usepackage{microtype}      % microtypography
\usepackage{fancyhdr}       % header
\usepackage{graphicx}       % graphics
\usepackage{subcaption}
\graphicspath{{media/}}     % organize your images and other figures under media/ folder

\usepackage{amssymb}
\usepackage{amsmath}
\usepackage{longtable}
\usepackage{lscape}
\usepackage{siunitx}
\usepackage{cleveref}
\usepackage{multirow}
\usepackage{xcolor}
\usepackage{color}
\usepackage{xspace}
\usepackage{url}
\newcommand{\systemname}{DuetML\xspace}

\usepackage[most]{tcolorbox}
\usepackage{varwidth}
\definecolor{block-gray}{gray}{0.97}
\usepackage{soul}
\newtcolorbox{blockquote}{
  colback=block-gray,
  grow to right by=-1mm,
  grow to left by=-1mm,
  boxrule=0pt,
  boxsep=0pt,
  breakable
}
\newcommand{\ctext}[3][RGB]{%
  \begingroup
  \sethlcolor{block-gray}% 新しく定義した色を設定
  \hl{#2}% テキストをハイライト
  \endgroup
}
\newif\ifdraft
\draftfalse

\ifdraft
    \newcommand{\key}[1]{\textcolor{red}{{Key: #1}}}
    
    \newcommand{\todo}[1]{\textcolor{red}{\textbf{TODO: #1}}}
    \newcommand{\sugano}[1]{\textcolor{orange}{{[Sugano: #1]}}}
    \newcommand{\kawabe}[1]{\textcolor{magenta}{{[Kawabe: #1]}}}
\else
    \newcommand{\key}[1]{}
    
    \newcommand{\todo}[1]{}
    \newcommand{\sugano}[1]{}
    \newcommand{\kawabe}[1]{}
\fi
\title{\systemname: Human-LLM Collaborative Machine Learning Framework for Non-Expert Users}

\author{
  Wataru Kawabe and Yusuke Sugano \\
  Institute of Industrial Science \\
  The University of Tokyo \\
  Tokyo, Japan\\
  \texttt{\{wkawabe, sugano\}@iis.u-tokyo.ac.jp} \\
}
\begin{document}
\maketitle
\begin{abstract}
Machine learning (ML) models have significantly impacted various domains in our everyday lives. 
While large language models (LLMs) offer intuitive interfaces and versatility, task-specific ML models remain valuable for their efficiency and focused performance in specialized tasks. 
However, developing these models requires technical expertise, making it particularly challenging for non-expert users to customize them for their unique needs. 
Although interactive machine learning (IML) aims to democratize ML development through user-friendly interfaces, users struggle to translate their requirements into appropriate ML tasks. 
We propose human-LLM collaborative ML as a new paradigm bridging human-driven IML and machine-driven LLM approaches.
To realize this vision, we introduce \systemname, a framework that integrates multimodal LLMs (MLLMs) as interactive agents collaborating with users throughout the ML process.
Our system carefully balances MLLM capabilities with user agency by implementing both reactive and proactive interactions between users and MLLM agents.
Through a comparative user study, we demonstrate that \systemname enables non-expert users to define training data that better aligns with target tasks without increasing cognitive load, while offering opportunities for deeper engagement with ML task formulation.
\end{abstract}

\begin{figure}
    \centerline{\includegraphics[width=.85\textwidth]{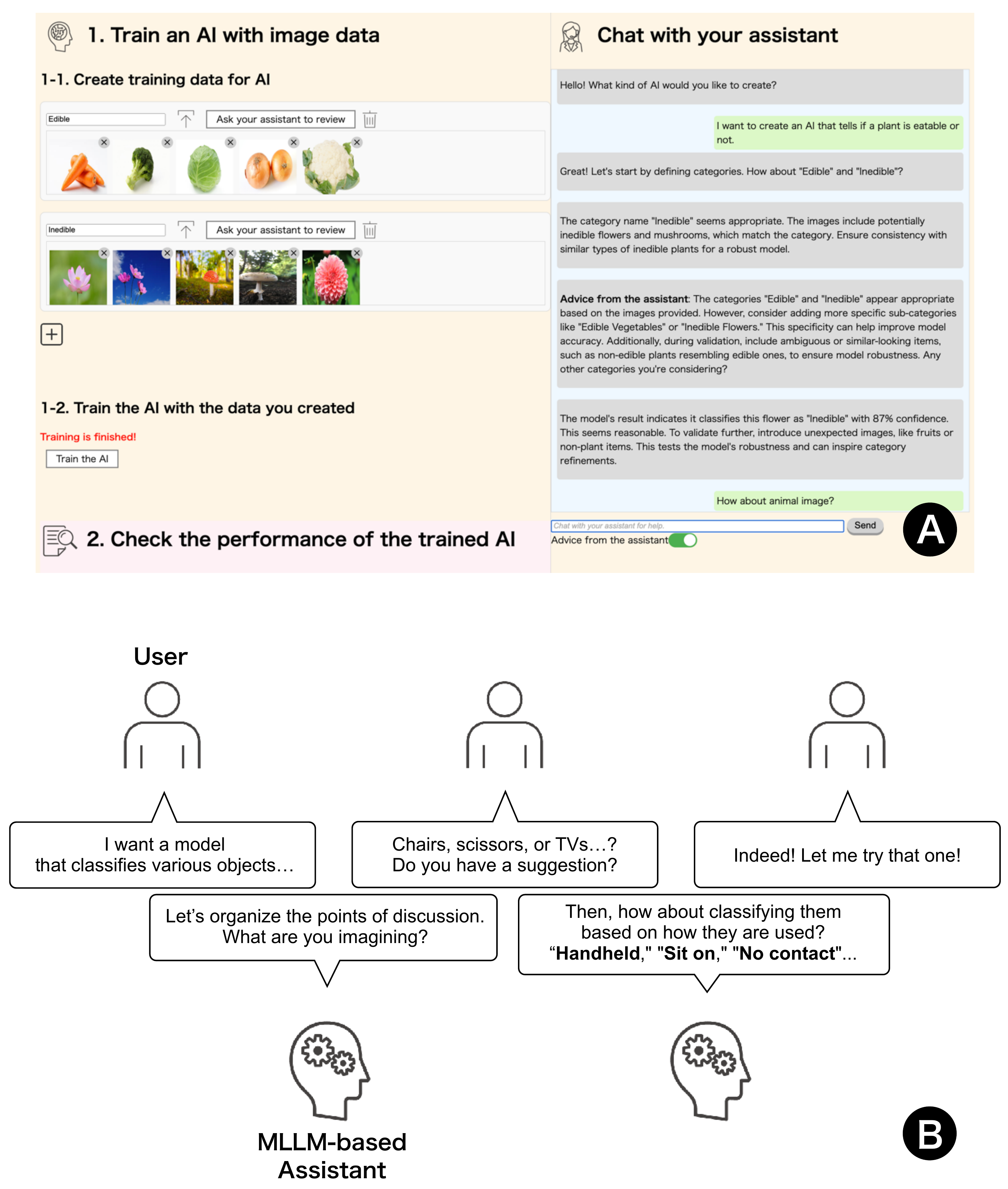}}
    \caption{(A) Our system \systemname aims to assist users without a technical background in appropriately formulating tasks and creating training data in machine learning prototyping. (B) The system uses a multimodal large language model (MLLM)-based assistant to elicit user needs and guide them interactively toward appropriate training data.}
    \label{fig:teaser}
\end{figure}


\section{Introduction}

% 段落1：MLモデルとユーザの背景
Machine learning (ML) models have found applications across various domains, significantly impacting multiple facets of modern society. 
ML approaches have evolved in performance over the years, from specialized task-specific models such as SVMs~\cite{hearst1998support} and convolutional neural networks (CNNs)~\cite{gu2018recent} to more recent general-purpose large language models (LLMs).
While LLMs have demonstrated remarkable capabilities in handling diverse tasks and offer intuitive natural language interfaces, they often require substantial computational resources. 
Furthermore, they may not excel in specific tasks such as object detection\cite{zhang2024good} or image classification~\cite{zhang2024visually}, and present notable challenges for non-expert users in crafting effective prompts~\cite{liu2023pre,kim2024evallm}. 
Task-specific ML models, including CNNs, remain valuable for their efficiency and effectiveness in specialized tasks, offering lightweight solutions with focused performance. 
However, they require task-specific training and struggle with unfamiliar domains. 
Additionally, they demand users to possess programming skills and mathematical knowledge for customization, making it particularly challenging for non-expert users to tune these models effectively for their unique needs.

% While LLMs have demonstrated remarkable capabilities in handling diverse tasks intelligently, they often require substantial computational resources and may not excel in specific tasks such as object detection~\cite{zhang2024good} or image classification~\cite{zhang2024visually}.
% Classical ML models and CNNs remain valuable for their efficiency and effectiveness in specialized tasks, offering lightweight solutions with focused performance.
% However, these models present challenges: they require task-specific training and struggle with unfamiliar domains.
% Users must leverage programming skills and mathematical knowledge to customize these models for specific tasks.
% This customization process is particularly challenging for non-expert users who lack technical backgrounds, making it difficult for them to tune models effectively for their unique needs.

% 段落2：IMLというのがある。ユーザはうまくやれない
To address these challenges and facilitate easier task-specific ML model development for non-expert users, interactive ML (IML) has been investigated as another paradigm~\cite{dudley2018review,amershi2014power}. 
Various IML systems are designed to enable users without a technical background to participate in ML development through user-friendly UI operations and to make ML more accessible to a broader audience. 
% The aim is to democratize ML by making it more accessible to a broader audience. 
Nevertheless, it has been reported that employing IML to develop models that appropriately solve tasks is not always a straightforward process for users~\cite{kawabe2024image,nakao2020use}. 
For example, users are challenged to translate their needs into appropriate tasks effectively. 
Suppose a scenario where users aim to create a model that identifies edible plants. 
The optimal approach involves using broader categories like ``Edible'', ``Non-Edible,'' and ``Non-Plant,'' rather than specific labels such as ``Tomato'' or ``Mushroom.''
This level of abstraction often proves difficult for novice users to grasp independently. 
This complexity indicates the need for further refinement and support to make the process intuitive for non-experts.

% 段落3（新規書き換え）：IMLの最中アドバイスが必要。それはMLLMの仕事かも
% In the context of IML systems for ML development, users face the challenge of effectively translating their needs into appropriate tasks. 
% Consider a scenario where users aim to create a model that identifies edible plants. 
% The optimal approach involves using broader categories like ``Edible'', ``Non-Edible,'' and ``Non-Plant,'' rather than specific labels such as ``Tomato'' or ``Mushroom.''
% This level of abstraction often proves difficult for novice users to grasp independently. 

% Providing objective guidance to users can be a promising approach to helping them formulate ML problems correctly and develop appropriate models for their needs. 
Building upon these observations, we propose \textit{human-LLM collaborative ML} as a new paradigm for task-specific ML model development.
As discussed above, IML and LLM can be seen as contrasting in terms of whether the initiative in task formulation is placed with humans or machines. 
Both the IML aiming for full control by human intelligence and the LLM aiming for the generalization of machine intelligence have their pros and cons when it comes to empowering non-expert users. 
We hypothesize that a process in which humans and machines collaboratively create task-specific ML models might be an alternative solution for non-expert users. 
Human-LLM collaborative ML seeks to fully leverage the expertise of LLMs by having them actively intervene in the formulation, while keeping humans as the primary agents in task setting.
From a user's perspective, this can also be viewed as a process where humans refine the diverse knowledge of LLMs into a task-specific model. 
On the other hand, compared to customizing the behavior of LLMs using only prompts, the example-based IML-style training process can be conducted more intuitively.
% hypothesizing that collaborative discussions could yield better results than solitary efforts. 
% State-of-the-art multimodal large language models (MLLMs) are promising collaboration partners for this paradigm, as they can comprehend subtle nuances from both text and images to provide intelligent responses. 
% However, the ML model development process involves complex steps, and complete MLLM automation remains challenging.
% Moreover, non-expert users often struggle to effectively communicate their intentions to MLLMs through prompts~\cite{liu2023pre,kim2024evallm}. 
% As a concrete framework within this paradigm, our research focuses on creating a system enabling productive human-MLLM collaboration throughout development.
% We design a system to facilitate cooperation between users and MLLMs, fostering mutual knowledge exchange to generate appropriate training data.

As a prototype to realize this paradigm, we introduce \systemname, a GUI-based interactive framework that promotes effective human-LLM collaborative ML by leveraging conversational capabilities of state-of-the-art multimodal large language models (MLLMs) (Fig.~\ref{fig:teaser}). 
Our system facilitates collaboration between users and MLLM-based agents, where the agents support the IML process by helping translate vague requirements into concrete ML tasks while preserving user autonomy.
% Our system ensures collaboration between the user and LLM, in which the MLLM-based agents cooperate with the user's IML process by translating vague requirements into concrete ML tasks while respecting their unique perspectives. 
Our \systemname is based on an IML-style UI, while the MLLM-based agents observe user interaction history, task descriptions, and training data definitions to assist and intervene in the user's work.
The user aims to train a lightweight classification model through interactive supervised learning in collaboration with MLLM agents.
% The agents maintain awareness of the user's context by analyzing training data, dialogue history, and other relevant information. 
Recognizing that optimal intervention styles vary among users, we employ two distinct agents: one responds reactively to user requests, while the other proactively offers suggestions about potentially overlooked aspects.
This dynamic partnership helps transform users' initial, often ambiguous ideas into appropriate training data through contextually aware interactions.
We conducted a study with non-expert users to evaluate the effectiveness of human-MLLM collaboration in ML task formulation.
Using a between-subjects design, we compared \systemname to a baseline IML system without collaborative features. 
Participants followed predefined instructions to create image classification models using their assigned system. 
The results showed that participants working collaboratively with the agents through \systemname created training data that better aligned with the target task.
While participants did not subjectively perceive \systemname as more usable than the baseline system, third-party evaluation of their defined category names revealed that \systemname led to superior results.
Additionally, subjective evaluations of \systemname's collaborative features revealed that participants found the interaction with the agents beneficial rather than cognitively demanding.
The MLLM's image recognition capability also enabled a direct understanding of users' training and inference data, allowing for more comprehensive guidance compared to text-only systems.
This suggests that human-MLLM collaboration can provide diverse support beyond task formulation, adapting to users' varying levels of technical understanding.
While our study results do not suggest that \systemname was universally beneficial across all aspects of the ML process, we found that it provided positive contributions, such as offering non-expert users opportunities to think more deeply about their ML tasks.
These findings not only demonstrate the potential of human-LLM collaborative ML for democratizing ML development but also open new avenues for discussion on how to balance human agency and machine intelligence in ML development.

\section{Related Work}

\subsection{Interactive Machine Learning for Non-Expert Users}

The major mission of IML research is to enable users without an AI/ML background to develop ML models without requiring technical knowledge~\cite{dudley2018review,fails2003interactive,amershi2014power}. 
To achieve this goal, researchers have approached the challenge from various perspectives.
For instance, researchers have created guidelines for IML that address the technical challenges non-expert users are likely to encounter~\cite{yang2018grounding,mishra2021designing}. 
Specifically, Yang et al.~\cite{yang2018grounding} researched how non-experts build their own ML solutions, revealing unique potentials and pitfalls, and proposed design implications to help non-experts create robust solutions easily. 
Building on this work, Mishra et al.~\cite{mishra2021designing} examined how non-experts use transfer learning in an interactive environment by designing prototypes, uncovering data- and perception-driven strategies, and proposing design implications to guide future interactive transfer learning tools. 
Another significant stream of research has focused on visualization approaches, where data embedded in a two-dimensional plane is presented to users for interaction~\cite{tatsuya2020investigating,chang2021spatial,kawabe2024technical}, and the performance of models is intuitively visualized~\cite{talbot2009ensemblematrix,ren2016squares,zhang2018manifold}.

However, despite these advances in making ML development more accessible, users without AI/ML literacy often struggle at an earlier stage: accurately formulating their needs into specific tasks.
While traditional IML approaches have focused on making the technical aspects of ML more approachable, language-based approaches could help solve this fundamental problem through user dialogue.
Yet, there is limited discussion on language-based solutions, and merely implementing a chat interface with LLMs proves insufficient for effective user support.
This is because, to provide meaningful intervention, LLMs must comprehend both the users' cognitive processes and their current contextual states. 
\systemname addresses this challenge by facilitating human-MLLM collaboration through mutual knowledge exchange between users and MLLM-based agents on the GUI, creating a bridge between traditional IML approaches and advanced language model capabilities.

\subsection{Human-LLM Interaction for Complex Tasks}

% HCI的な例で、AI Assistant/copilot/collaboratorの助言を一気に攫う

In the HCI field, the context of a computer assistant or intelligent agent collaborating with users through intervention for more efficient and effective technical interactions has been studied. 
As a fundamental framework in this field, Horvitz defined principles for mixed-initiative UI, where automated services from the computer and direct manipulation from the human cooperate to achieve the goal~\cite{horvitz1999principles}. 
Following this direction, design guidelines have been proposed for how intelligent assistants/copilots/collaborators can recognize user states and meet their expectations through interaction~\cite{rapp2021human,alkatheiri2022artificial}. 
The assistant/copilot/collaborator is expected to resolve users' questions and enable them to utilize their capabilities fully. 
However, while researchers have discussed guidelines and approaches for intelligent assistance, previous intelligent systems lacked sufficient capabilities in context understanding and language expression, such as generating natural responses and understanding user intent. 
This limitation has prevented the development of interactive systems that fully achieve the ideal state.

% % AIとかいう以前に、コーチングのインタラクションにはHCIではどういう方法が可能か、これまでどういう文脈だったか、コーチングには何が期待されているか、という大きな話
% In the HCI field, the context of a computer assistant or intelligent agent collaborating with users through intervention for more efficient and effective technical interactions has been studied.
% Horvitz defined principles for mixed-initiative UI, where automated services from the computer and direct manipulation from the human cooperate to achieve the goal~\cite{horvitz1999principles}.
% More specifically, design guidelines have been proposed for how intelligent assistants/copilots/collaborators can recognize user states and meet their expectations through interaction~\cite{rapp2021human,alkatheiri2022artificial}.
% The assistant/copilot/collaborator is expected to resolve users' questions and enable them to utilize their capabilities fully. 
% While researchers have discussed guidelines and approaches for intelligent assistance, previous intelligent systems lacked sufficient capabilities in context understanding and language expression, such as generating natural responses and understanding user intent. 
% This limitation has prevented the development of interactive systems that fully achieve the ideal state.

% ユーザのインタラクションを支援するためにAIが助言を与えるケース。コーディング~\cite{ross2023programmer}やMR~\cite{bohus2024sigma,pei2024autonomous}、視線インタラクション~\cite{}の例
With recent advances in language-based models like LLMs, monitoring user interactions and collaborating with them has been applied to various application scenarios.
In the context of creative assistance, e.g. programming, it is possible for an assistant to directly refer to the user's code and give advice~\cite{ross2023programmer,pinto2024lessons,kazemitabaar2024codeaid}.
The LLM-based collaborators are useful not only in programming but also in creative activities in general, such as document creation~\cite{levine2024students,salvagno2023can} and artistic design~\cite{guo2023ai,sano2022ai}.
In the mixed reality, the intelligent advisors recognize the surrounding environment to advise on the user's physical interactions~\cite{bohus2024sigma,pei2024autonomous}.
The interactive systems with a collaborative intelligent agent are being explored in educational scenarios to serve as tutors that monitor students' learning progress and provide advice based on it~\cite{chen2020artificial,hicke2023chata}. 
Similarly, in health and fitness contexts, an LLM-based teacher is expected to play the role of a personalized advisor~\cite{nahavandi2022application,ong2024advancing,jorke2024supporting,meywirth2024designing}. 
Numerous scenarios exist where an assistant/copilot/collaborator can intervene in user activities, and with the advent of LLMs and/or MLLMs, the accuracy of providing context-aware advice has improved dramatically.
These technological advances open new possibilities for applying MLLMs in specialized domains such as ML model development.

% \key{我々の研究は、小さなMLモデルを作る過程そのものにエージェントを導入する。これにより、エージェントがユーザの意図、教師データ、モデルのパフォーマンスといったIMLプロセスに含まれる動作全体を俯瞰し、次に行うべきユーザのアクションを提案することができる}
Building upon these advances in context-aware assistance through human-LLM interaction, we propose human-MLLM collaborative ML as a new paradigm, where MLLM-based agents intervene in and cooperate with users' interaction processes as one implementation framework.
The agents can comprehensively oversee users' thought processes by analyzing the history of user-agent dialogues and the resulting training data created by users.
These agents can suggest users' next actions and serve as discussion partners when users formulate their challenges into ML tasks.
This approach leverages the (M)LLM's ability to provide context-aware recommendations or guidance, streamlining the iterative ML model refinement or customization process.
% Our study proposes a novel framework for human-MLLM collaborative ML by introducing MLLM-based agents to users developing and customizing their models. 
% This integration enables the agents to comprehensively oversee the users' thought process, including managing the training data and evaluating the model's performance. 
% By doing so, the agents can effectively suggest the next actions the user should take. 
% This approach leverages the (M)LLM's ability to provide context-aware recommendations or guidance, streamlining the iterative ML model refinement or customization process.

\subsection{Interaction Methods of MLLMs}

% もうちょい技術的な、LLM研究自体の話

% MLLMs are well suited for intelligent interactions because of their broad general knowledge and fluent language outputs~\cite{wu2023next,xu2024survey,xie2024large}.
% The term ``multimodal'' refers to the ability to input and output modalities other than text, with images being an primary example. 
% Major models like GPT-4~\cite{achiam2023gpt} and Gemini 1.5~\cite{reid2024gemini} are already designed to be multimodal, enabling broader and more diverse understanding through image recognition compared to text-only interactions. 
% % allowing for a detailed understanding of the context in which an image is situated when combined with optical character recognition~\cite{memon2020handwritten,islam2017survey}. 
% MLLMs' image recognition capabilities have enabled rich contextual understanding, such as medical image diagnosis~\cite{panagoulias2024evaluating} and emotion recognition~\cite{kumar2024multimodal}. 
% Considering ML contexts, we considered that user-generated data, such as training or test data, could serve as direct indicators of user states. 
% We utilize MLLM's image recognition capabilities when incorporating user-defined training or test data.

MLLMs are well suited for intelligent interactions because of their broad general knowledge and fluent language outputs~\cite{wu2023next,xu2024survey,xie2024large}. 
The term ``multimodal'' refers to the ability to input and output modalities other than text, with images being a primary example. 
Major models like GPT-4~\cite{achiam2023gpt} and Gemini 1.5~\cite{reid2024gemini} are already designed to be multimodal, enabling broader and more diverse understanding through image recognition compared to text-only interactions. 
This capability has proven valuable in various domains, enabling rich contextual understanding in tasks such as medical image diagnosis~\cite{panagoulias2024evaluating} and emotion recognition~\cite{kumar2024multimodal}. 
These successful applications of MLLMs in context-dependent tasks suggest their potential value in ML development contexts, where user-generated data, such as training or test data, could serve as direct indicators of user states. 
Based on this potential, we utilize MLLM's image recognition capabilities when incorporating user-defined training or test data.

% \key{LLMの技術的な話。Conversational Interactionへの活用の仕方とか、プロンプトの作り方とか、モデルによってできることの違いとか。あとはRAGみたいなLLM同士をいかに繋げていろんなタスクを溶かせるかみたいな話とか。}
% % There are several methods for an MLLM to understand the unique circumstances of a user. 
% Several methods enable MLLMs to generate outputs customized to each user's specific situation and data.
% % For example, retrieval-augmented generation~\cite{lewis2020retrieval,xu2024retrieval,shuster2021retrieval} allows an MLLM model to access and reference external knowledge by searching unique information sources. 
% The most naive approach is to fine-tune (M)LLMs with user-provided data. 
% Various efficient training methods have been proposed to adapt (M)LLMs to user-specific contexts~\cite{tan2024democratizing,lin2024data,kuang2024federatedscope}.
% Additionally, even without fine-tuning, there are cases where MLLMs can generate outputs while referencing user-dependent contexts with each new input by incorporating interaction history and specific knowledge~\cite{sanchez2024automating,maharana2024evaluating,baek2024knowledge}.
% In practice, when interacting LLM-based agents with each other, each agent maintains an understanding of the context by inputting the history of the situation~\cite{park2023generative}.
% Similarly, in \systemname, the MLLM agents reference the chat history and the resulting training data that has been created so far. 
% This allows the agents to provide advice to the user with as complete an understanding of the user's state as possible.

Several methods enable MLLMs to generate outputs customized to each user's specific situation and data. 
One fundamental approach is to fine-tune (M)LLMs with user-provided data, and various efficient training methods have been proposed to adapt (M)LLMs to user-specific contexts~\cite{tan2024democratizing,lin2024data,kuang2024federatedscope}. 
However, even without such model modifications, MLLMs can generate context-aware outputs by incorporating interaction history and specific knowledge with each new input~\cite{sanchez2024automating,maharana2024evaluating,baek2024knowledge}.
This context-aware approach has been successfully demonstrated in multi-agent scenarios, where LLM-based agents maintain situation understanding by referencing interaction history~\cite{park2023generative}. 
Drawing from these insights, in \systemname, the MLLM agents reference both the chat history and the resulting training data that has been created so far, enabling them to provide advice with a comprehensive understanding of the user's state.

\section{\systemname}

\subsection{System Design Concept}

% IMLシステムを用いても、初心者はタスクをうまく解けないことが知られている
% 我々は、MLLMの持つ視覚的、言語的に状況を理解してユーザと協働するという性質が、ここに役に立つのではないかと考え
% 一方で、MLLMをそのまま使うだけでは、MLLMは軽量MLモデル開発に適してはいないので不便である
% そこで我々は、従来のIMLシステムでできることは尊重しつつ、そこにMLLMをアシスタントとして入れ込むことでユーザがあいまいなアイデアを具体化していく作業を協調的に手伝うという形式を採用した
% システムではユーザはIMLを行うが、そのプロセスをアシスタントが監視しており、いくつかの方法でユーザに働きかけることで一緒に課題の解き方を具体化していく
To address the challenges non-expert users face in IML systems, we developed \systemname, which introduces MLLMs as collaborative agents in the ML development process.
When non-expert users interact with an IML system to proceed with ML development, they sometimes struggle to formulate tasks effectively and create training data~\cite{kawabe2024image,nakao2020use}.  
We hypothesized that MLLMs could mitigate this issue through their ability to comprehensively understand users' operational status visually and linguistically, enabling contextual collaboration with users.
However, directly using MLLMs might limit interactions to users merely extracting general knowledge, potentially leading to misaligned communication where MLLMs fail to understand and address users' specific needs properly.
To enable meaningful interactive communication between users and MLLMs, we designed a framework that incorporates MLLMs as collaborative agents, enabling interactive model training and evaluation through conversational interaction with the user.
To realize this vision, we carefully designed how the MLLM agents should participate in the user's ML process.

% MLLMはユーザのインタラクションプロセスを全て把握している。干渉の仕方はユーザに合わせていくつかのパターンを用意している。
In \systemname, as users progress through the stages of ML development, including creating training data and training models and evaluating the trained models, the MLLM agent accompanies them to help concretize their ideas. 
% Users interact with the agent via a chat interface, and it continuously monitors and comprehends all prior interactions, including chats.
% We believe the appropriate way for the agent to intervene varies from user to user. 
Users interact with the agent via a chat interface, and it continuously monitors and comprehends all prior interactions, including chats. 
Based on this comprehensive understanding of user interactions, we believe the appropriate way for the agent to intervene varies from user to user.
Therefore, we designed a system that operates two agents simultaneously, each engaging with users at different levels of proactivity.
These agents, termed \textbf{active} and \textbf{passive agents}, serve complementary roles.
Users can proactively seek ideas and guidance from the active agent, while the passive agent automatically generates advice to point out aspects that users might overlook, even without explicit user requests.

\begin{figure}[t]
\centerline{\includegraphics[width=.7\textwidth]{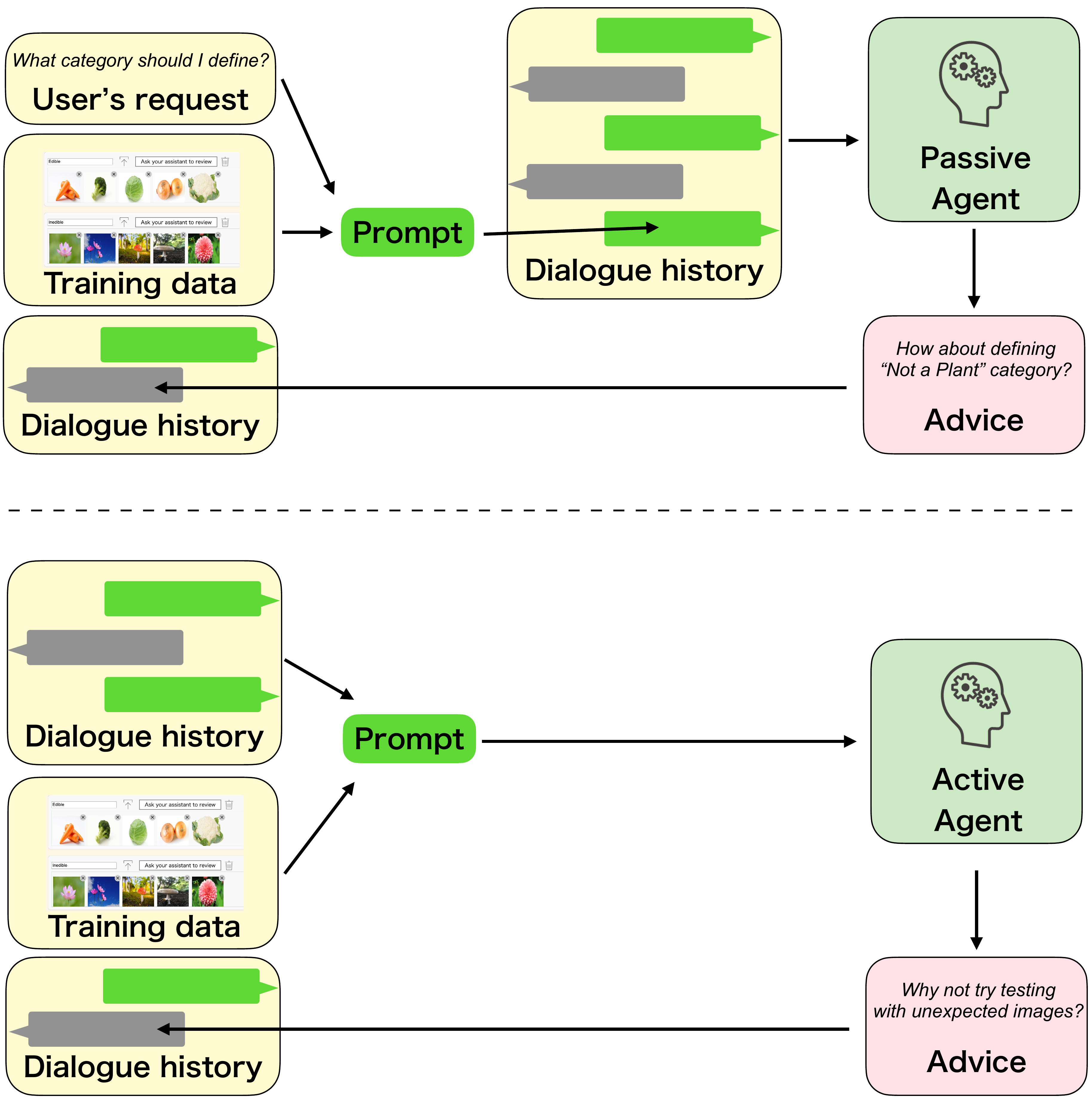}}
\caption{Each agent's data processing flow in \systemname. The passive agent receives the dialogue history, including the latest prompt about the user's request and training data. The active agent receives dialogue history and the training data. The generated advice is appended to the dialogue history.}
\label{fig:flow_imla}
\end{figure}

% まずMLLM設計の概要
The technical core of \systemname lies in its approach of providing MLLMs with context that includes visual information and user interaction states.
To implement this approach, the agents receive prompt information about the user's interaction context, including dialogue history and training data images. 
These carefully designed prompts reflect the core philosophy of enabling the agents to collaborate with the user by providing advice based on a comprehensive understanding of the user interaction.
This is reflected in the prompts with text that states: ``\ctext{\texttt{Support the user in formulating their tasks by proposing specific examples through dialogue.}}'''
This guides the agents in providing targeted and contextually relevant support. 
To realize this interaction mechanism, we implemented two distinct processing flows for the passive and active agents, as shown in Fig.~\ref{fig:flow_imla}. 
The passive agent responds to explicit user requests, following a straightforward interaction pattern where user requests and training data are embedded in a prompt template and sent to the passive agent as the latest user utterance in the dialogue history. 
In contrast, the active agent generates advice proactively by referencing all available information rather than waiting for user prompts.
This agent receives prompts containing the dialogue history and training data to provide context-aware suggestions. 
In both cases, generated advice is appended to the end of the dialogue history. 
Detailed prompt specifications for each agent can be found in the Appendix.
% Figure~\ref{fig:flow_imla} shows the data processing flow for each agent.
% The passive agent responds to explicit user requests. 
% User requests and training data are embedded in a prompt template and sent to the passive agent as the latest user utterance in the dialogue history.
% In contrast, the active agent generates advice proactively by referencing all available information rather than waiting for user prompts. 
% This agent receives prompts containing both the dialogue history and training data to provide context-aware suggestions.
% In both cases, generated advice is appended to the end of the dialogue history.
% Detailed prompt specifications for each agent can be found in the Appendix.

\subsection{Interaction Flow}

\begin{figure}
    \centerline{\includegraphics[width=.9\textwidth]{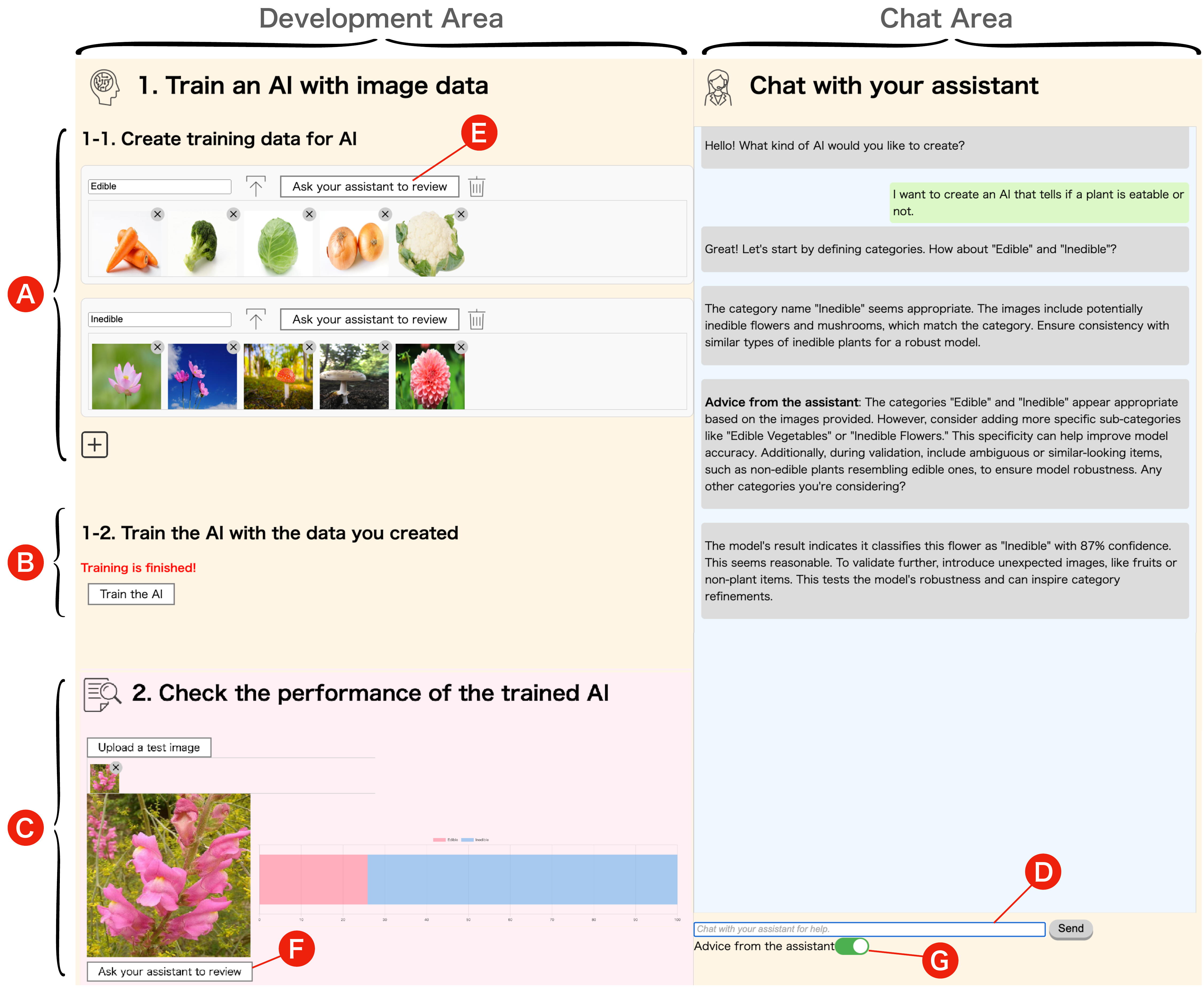}}
    \caption{The overview of \systemname. Users can create training data for the classification model (A), train the model with it (B), and evaluate the trained model's performance (C) in the IML area. During this IML process, the passive agent provides advice in response to either the user's chat input (D) or button inputs related to training data (E) or inference results (F). Additionally, the active agent monitors the user's overall interaction and periodically offers advice. Users can toggle this feature on or off as they prefer (G). All the advice from the agents is shown in the chat area.}
    \label{fig:proposed_imla}
\end{figure}

% ユーザはIMLプロセスを進めながらアシスタントともインタラクションする
% Figure~\ref{fig:proposed_imla} illustrates the GUI of \systemname. 
% The interface is primarily divided into two areas: the development area and the chat area.
% In the development area, users engage in the fundamental ML development process: defining training data (A), model training (B), and performance evaluation (C). 
% The MLLM-based agents monitor the entire process and guide via the chat interface throughout this workflow.
Figure~\ref{fig:proposed_imla} illustrates the GUI of \systemname. 
The interface is primarily divided into two areas: the development area and the chat area.
In the development area, users engage in the fundamental ML development process: defining training data (A), model training (B), and performance evaluation (C). 
The MLLM-based agents monitor the entire process and guide via the chat interface throughout this workflow.
To facilitate effective interaction between users and agents, we implemented two distinct interaction patterns that complement each other.

The first interaction pattern involves the passive agent, which responds to users' explicit requests for assistance.
Users can actively seek advice when they have questions. 
When users start using the system, the agent first asks ``\ctext{\texttt{What kind of AI would you like to create?}}''' 
Users must respond with their goal through chat input (D), such as ``\ctext{\texttt{I want to create a model that can determine if plants are edible.}}''' 
Users and the system must conduct this interaction every time users start using it, allowing the agents to grasp their vague needs. 
The interaction continues from there, with users asking questions through the same chat input, like ``\ctext{\texttt{What kind of training data should I prepare?}}''' for example. 
Additionally, the ``Ask the Assistant'' buttons are strategically placed in the training data creation (E) and model evaluation sections (F), enabling users to request targeted advice for these specific tasks. 
User actions such as chat inputs and ``Ask'' button clicks are converted into the user's request and included in the prompt sent to the agent (see Fig.~\ref{fig:flow_imla}).

The second interaction pattern is implemented through the active agent, which provides proactive support. 
This agent monitors the interaction for users who are not actively seeking advice and periodically offers suggestions to guide them effectively. 
For instance, it might propose, ``\ctext{\texttt{Consider using broader category names like `Edible Plants' and `Non-Edible Plants' instead of specific names like `Carrot' and `Flower'.}}''' 
Users have the option to turn off this agent using a toggle button if they think it is unnecessary (G). The active agent receives prompts containing dialogue history and current training data as indicators to understand the user's state.

\subsection{Use Case Scenario}
\begin{figure*}
    \centerline{\includegraphics[width=.8\textwidth]{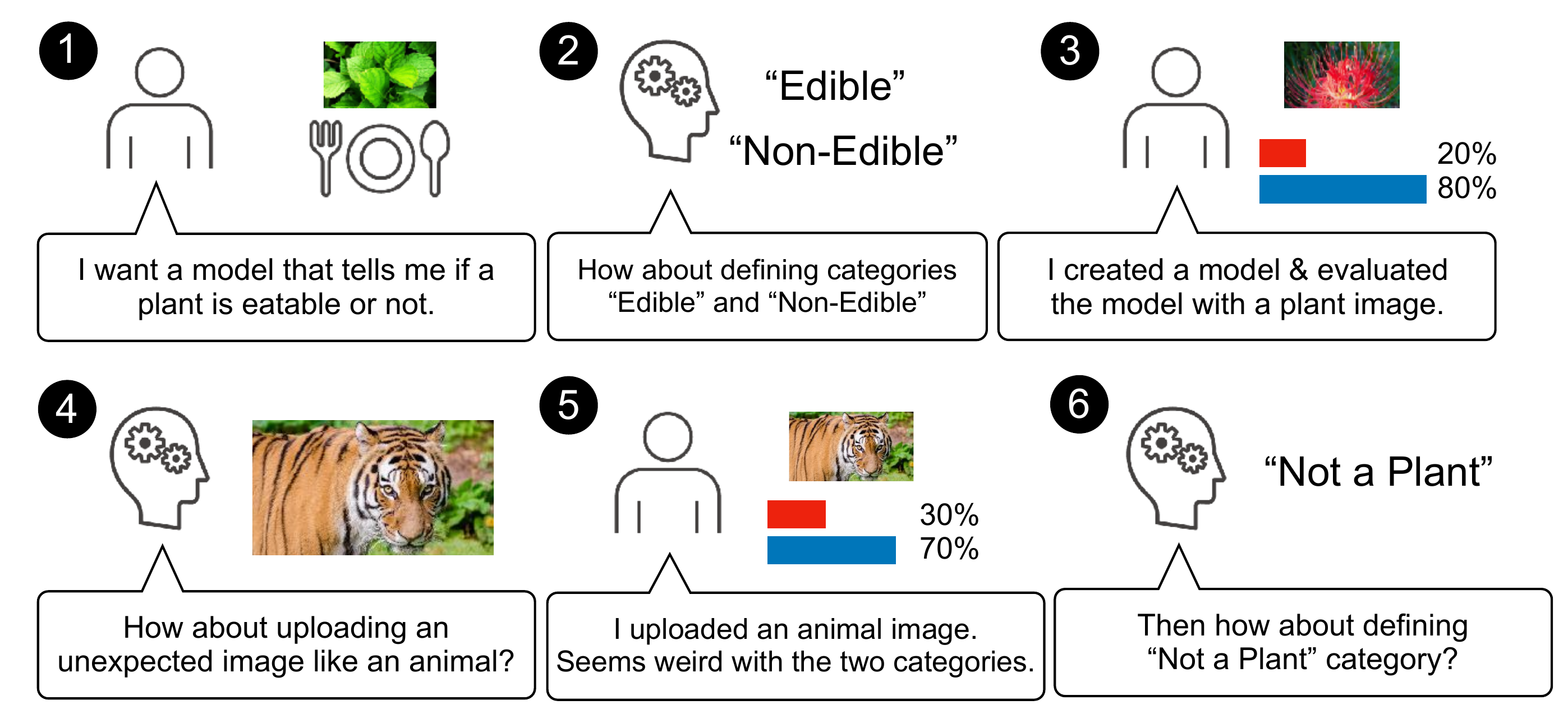}}
    \caption{An example use case scenario of \systemname.}
    \label{fig:use_case_imla}
\end{figure*}

We illustrate an example use case in Fig.~\ref{fig:use_case_imla}. 
Consider a user aiming to develop a model that identifies edible plants in images. 
At the beginning of the development process, the user, uncertain about the category definition, expresses this requirement via chat.
Recognizing the user's uncertainty, the passive agent proposes, ``\ctext{\texttt{How about defining two categories: `Edible' and `Non-Edible'?}}''' 
Following this guidance, the user adopts this suggestion and begins populating the `Edible' category with vegetable images and the `Non-Edible' category with wild plant images.
To ensure the appropriateness of their approach, the user utilizes the ``Ask'' button to seek confirmation on the suitability of these images as training data. 
With positive feedback from the passive agent, the user proceeds with model training.

As the development process moves into the evaluation phase, the user tests the trained model using various plant images.
During this evaluation, when the user tests non-edible flower images, the active agent proactively identifies a potential limitation in the current approach and advises: ``\ctext{\texttt{To enhance accuracy, consider introducing unexpected categories like fruit or animals.}}'''
The significance of this suggestion becomes apparent when the model, limited to the `Edible' and `Non-Edible' categories, produces inconsistent results with an animal image.
Recognizing this issue, the user queries: ``\ctext{\texttt{It seems odd to classify a lion image as `Edible' or `Non-Edible.'}}'''
To address this concern, the passive agent recommends creating a new ``Not a Plant'' category. 
The user implements this advice through this iterative refinement process by adding animal images to the new category and retrains the model. 

This example demonstrates how our system facilitates an interactive learning process in which users can progressively refine their approach based on both proactive and reactive guidance from agents.
Unlike conventional MLLM-based applications like chatbots or copilots, our system offers users insights that transcend mere assistance by enabling them to discover and address potential limitations in their initial approach.

\subsection{Implementation Details}

Our system implementation consists of three main components: the ML model architecture, the web-based interface, and the MLLM integration.
The classification model connects MobileNet~\cite{howard2017mobilenets} with a Support Vector Machine (SVM)~\cite{cortes1995support}, where the former extracts features and the latter performs inference using the extracted feature vectors.
During training, only the SVM is trained, and the parameters of MobileNet are fixed.
We adopted this architecture and the training method because the primary priorities in our user study are the simplicity and efficiency of the training process rather than the accuracy obtained through the training of state-of-the-art models, such as deep neural networks or LLMs. 
We used Scikit-learn's SVC implementation\footnote{\url{https://scikit-learn.org/dev/modules/generated/sklearn.svm.SVC.html}} as SVM and Torchvision's pre-trained MobileNet V2 model\footnote{\url{https://pytorch.org/hub/pytorch_vision_mobilenet_v2/}}.
Please note that our design can be applied to any classification model.

To enable interactive ML development, we implemented a web-based interface where the frontend is written in HTML and jQuery, and the backend is implemented in Python using Flask. 
The system handles training data management by sending it to the server through image files for each category, where images are arranged in a grid layout according to their categories.
For the user study, while all GUI components were translated into the native language of the experiment participants (Japanese), we used the original English prompts shown in the Appendix.

The MLLM integration is implemented through the OpenAI API (gpt-4o-2024-05-13 model)\footnote{\url{https://openai.com/index/hello-gpt-4o/}}, with different handling mechanisms for each agent type.
The passive agent's prompts are sent to the server in response to user requests, while the active agent operates on a regular 60-second interval.
To maintain system responsiveness, the prompt transmission is skipped if there are no user interactions during this interval.
Given that these operations are asynchronous and using a single MLLM would cause request conflicts, we implemented two separate GPT-4o models working simultaneously as individual agents.

\section{User Study}

To investigate whether \systemname can interactively support users in formulating ML tasks and how users perceive the experience of using it, we conducted a comparative user study. 
Since our interest lies in whether people without a technical background can accurately formulate ML problems, we targeted participants who are non-experts with no prior experience in AI/ML. 
These participants were tasked with creating training data, training a model with it, and evaluating the trained models to solve a predefined task provided by the researchers.

\begin{figure}
    \centerline{\includegraphics[width=.7\textwidth]{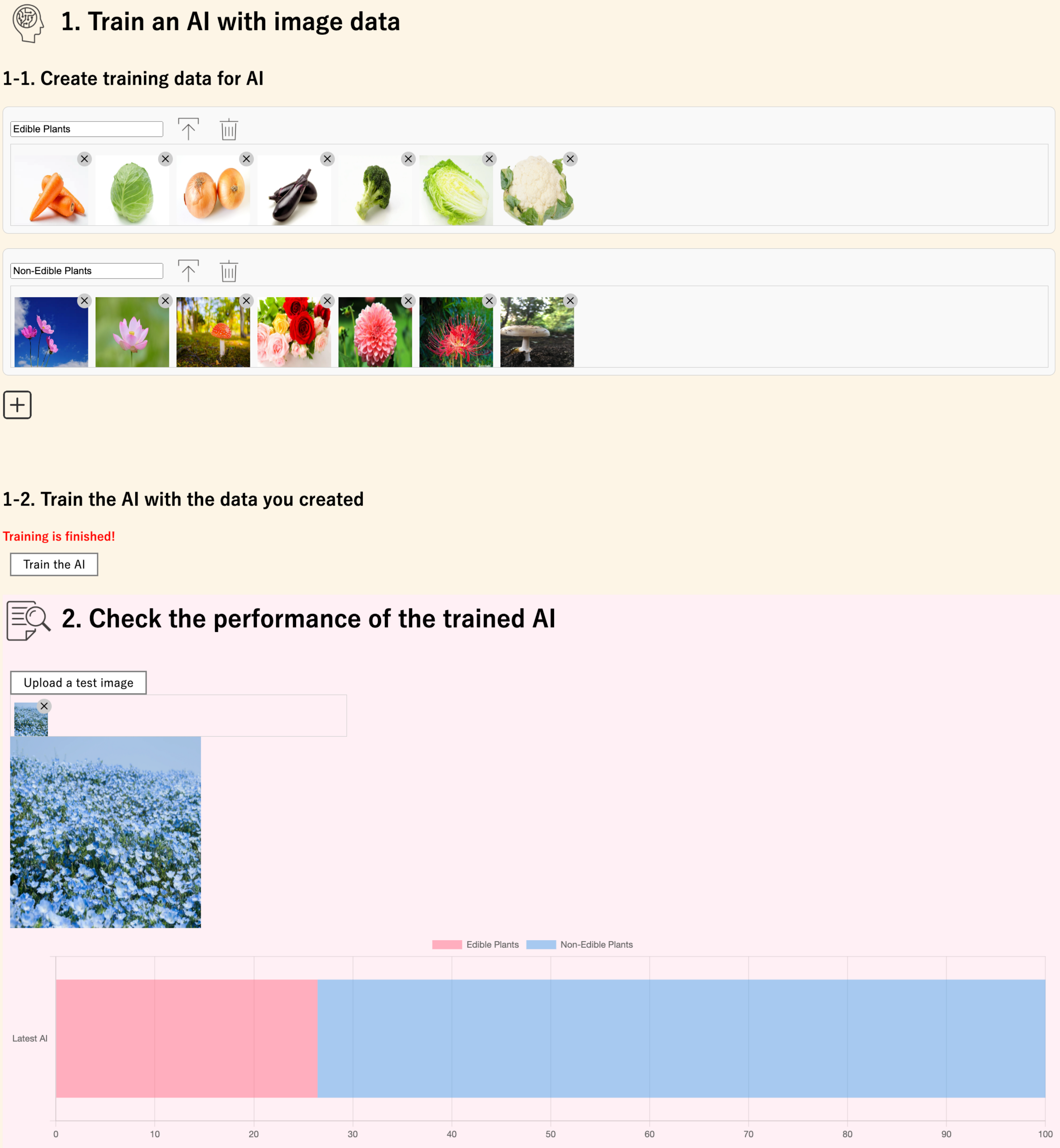}}
    \caption{The baseline system for the user study. Except for the absence of a chat area in \systemname, the UI design is consistent with \systemname.}
    \label{fig:baseline_imla}
\end{figure}

For comparison, we implemented a baseline system that provides the same ML development functionality without the agent features (GUI overview shown in Fig.~\ref{fig:baseline_imla}). 
To ensure fair comparison, we designed the baseline system's GUI to be as similar as possible to \systemname, and kept the backend components, including the ML model and training procedures, identical between the two systems. 
This design allowed us to isolate the impact of the agents' presence on users' task formulation process and overall experience.

% To investigate whether \systemname can interactively support users in formulating ML tasks and how users perceive the experience of using \systemname, we conducted a user study that compared \systemname with a simpler interactive system for ML development that excluded the agent functionality (hereafter referred to as the baseline system). 
% Since our interest lies in whether people without a technical background can accurately formulate ML problems, we targeted participants who are non-experts with no prior experience in AI/ML.
% Participants were tasked with creating training data, training models, and evaluating the trained models using either \systemname or the baseline system to solve a predefined task provided by the researchers.

% % ベースラインシステムの概要
% To examine the impact of the agents' presence, we implemented a baseline GUI-based system for ML development without the agent functionality as a comparison to \systemname (GUI overview shown in Fig.~\ref{fig:baseline_imla}). 
% In the baseline system, participants can create training data, train the model, and evaluate the trained model's performance.
% To minimize the influence of GUI differences on participants' perceptions, we designed the GUI of the baseline system to be as similar as possible to \systemname.
% The backend components, including the ML model and training procedures, were identical between the two systems.

\subsection{Procedure}

% % 参加者をどうやって集めたか、どうやって来てもらったか→説明まで
Before the study, we obtained approval from the university's ethics review committee. 
To recruit appropriate participants, we distributed announcements through university mailing lists and online bulletin boards, specifically requiring participants with ``no prior learning or implementation experience in ML'' to ensure only non-experts would participate. 
When participants arrived at the lab at their designated times, we provided an overview of the study task (i.e., creating an image classification model) and obtained their informed consent through a signed form.
Each participant was randomly assigned to either \systemname or the baseline system, with an equal number of participants in each condition.
To ensure consistent experimental conditions, we provided only essential operational instructions about the system. 
Following this brief introduction, participants were given a 15-minute familiarization period using a dataset of vegetable images~\cite{ahmed2021dcnn}, allowing them to freely explore the system's functionality.

% \key{こちらが与えた2つの課題に対してそれぞれ15分ずつ使ってシステムでモデルを作ってもらった}

The study consisted of two tasks, each lasting up to 15 minutes.
We instructed them to use the system for at least 10 minutes per task, after which they could stop at their discretion. 
The first task (the \textbf{open-ended task}) was designed to observe participants' natural approach to creating training data without specific constraints. 
Using a dataset containing various animal images\footnote{\url{https://www.kaggle.com/datasets/iamsouravbanerjee/animal-image-dataset-90-different-animals}}, participants were simply instructed to ``create an AI for classification using a collection of animal images.''
The second task (the \textbf{directed task}) examined participants' ability to meet specific classification requirements using the Caltech-101 dataset~\cite{fei2006one}. 
This task provided more structured guidance: ``Create a classification AI based on the provided images of objects, considering how these objects are used. 
For example, hammers and scissors are used by hand, and hats and shoes are worn on the body. 
Clocks and lamps function independently, and some objects are not man-made, such as living beings and food.''
Through subsequent third-party evaluation, we analyzed whether participants successfully created training data that met these task requirements.

To prevent potential biases, we preprocessed all datasets by shuffling images and renaming files with sequential numbers, regardless of their original categories.
Participants received written instructions for both tasks before beginning and were free to use any images from the provided folders. 
Throughout the study, we recorded system operation logs and chat logs with the agents for analysis. 
Upon completing both tasks, participants filled out a Google Form questionnaire and participated in an interview. 
The study concluded that participants received an Amazon Gift Card as compensation.

% We shuffled all images regardless of category to prevent bias from inherent groupings or hierarchical structures within the datasets. 
% We renamed the files with sequential numbering before providing them to participants. 
% Participants were free to use any images in the folder during the tasks.
% We provided written instructions for both tasks before participants began working on them. 
% The system operation logs and chat logs with the agents were recorded for our analysis. 
% After completing both tasks, participants answered a Google Form questionnaire and participated in an interview. 
% They then received an Amazon Gift Card as compensation, concluding the whole study.

% \subsection{Data Collection}

% Questionnaire
To evaluate specific aspects of participants' system experience, we assessed the system's usability using a 5-point Likert scale in the questionnaire. 
For both systems, we asked the following questions:
\begin{description}
    \item[Q1] You were able to work without technical difficulties.
    \item[Q2] You were finally able to create an AI model that meets the requirement.
    \item[Q3] The data you finally fed to the AI was appropriate.
    \item[Q4] You were able to work on the task with enjoyment.
    \item[Q5] You felt you could train an AI model yourself.
\end{description}
For participants who experienced \systemname, we included additional questions (Q6-Q10) to investigate their subjective experiences with the collaborative agents:
\begin{description}
    \item[Q6] You found the agent's advice accurate.
    \item[Q7] You found the agent's advice helpful.
    \item[Q8] You found the feature ``agent's tips'' useful.
    \item[Q9] You found the feature ``chat conversation'' useful.
    \item[Q10] You found the feature ``tell-me buttons'' useful.
\end{description}
In addition to the Likert scale items, we included an open-ended question: ``Please describe the system's good and bad points.''

% \kawabe{Describe about IMI?? <- Maybe no.}

% インタビュー
In the follow-up interviews, we asked each participant about the following for each task: (1) What kind of training data they initially planned to create, (2) What the final training data looked like, and (3) Whether there were any differences between the initial plan and the final result, and if so, what interactions with the system or what thoughts of the participants led to these changes.

% 第三者評価をする
\subsection{Third-Party Evaluation}

To objectively assess how participants could formulate tasks and create training data that meet the requirements of the directed task using both \systemname and the baseline system, we conducted a third-party evaluation of the participants' training data. 
This evaluation was necessary because determining whether training data meets specific task requirements cannot be measured through simple quantitative metrics alone.
Therefore, we designed an evaluation framework based on predetermined criteria and recruited experienced ML practitioners as evaluators.
To ensure evaluator expertise, we specified a requirement of ``at least two years of experience as an ML engineer or researcher in an ML-related field'' when recruiting through an online bulletin board. 
To maintain consistency in the evaluation process, all evaluators assessed the finalized training data from all study participants based on the following criteria:
\begin{description}
    \item[EQ1-a] (5-point Likert scale) Overall, category names express object usage. 
    \item[EQ1-b] (Free response) List IDs of categories where names express object usage. If none, write ``None.''
    \item[EQ2-a] (5-point Likert scale) Ignoring category names, images are generally classified based on object usage. 
    \item[EQ2-b] (Free response) List IDs of categories where images are classified based on object usage. If none, write "None."  
    \item[EQ3-a] (5-point Likert scale) Overall, category names and associated images match in content.
    \item[EQ3-b] (Free response) List IDs of categories where names and images match in content. If none, write ``None.''
    \item[EQ4-a] (5-point Likert scale) The categories comprehensively cover object usage methods.
    \item[EQ4-b] (Free response) Suggest missing category names for comprehensive coverage of object usage. If none, write ``None.''
    \item[EQ5-a] (5-point Likert scale) Categories are clearly distinguished without overlap or inclusion relationships.
    \item[EQ5-b] (Free response) List pairs of categories with overlap or inclusion relationships. If none, write ``None.''
\end{description}
% - (5段階のLikert scale) 全体的な傾向として、カテゴリ名は物体の使用方法を示す表現になっている
% - (自由記述) カテゴリ名が物体の使用方法を示す表現になっていると思うカテゴリのIDを挙げてください。なければ「なし」としてください。
% - (5段階のLikert scale) カテゴリ名を無視して画像だけを見たとき、全体的な傾向として、物体の使用方法に基づいて分類されている
% - (自由記述) 画像が物体の使用方法に基づいた分類になっていると思うカテゴリのIDを挙げてください。なければ「なし」としてください。
% - (5段階のLikert scale) 全体的な傾向として、カテゴリ名とそれに紐付けられた画像が内容的に一致している
% - (自由記述)カテゴリ名と画像が内容的に一致していると思うカテゴリのIDを挙げてください。なければ「なし」としてください。
% - (5段階のLikert scale) カテゴリの集合は、物体の使用方法を網羅的にカバーしている
% - (自由記述)物体の使用方法を網羅的にカバーする上で、欠けていると思うカテゴリ名をいくつか書いてください。なければ「なし」としてください。
% - (5段階のlikert scale)各カテゴリ同士は明確に区別され、重複や包含関係がない
% - (自由記述)重複、もしくは包含関係にあるカテゴリの組み合わせを挙げてください。なければ「なし」としてください。
To quantitatively compare the two systems, we assigned numerical values during analysis for Likert scale items: strongly disagree (-2), disagree (-1), neutral (0), agree (1), and strongly agree (2).

\subsection{Results}
We collected data from 12 participants (4 males and 8 females), ages 21 to 53 (M $= 34.83$, SD $= 9.74$).

\subsubsection{User Subjective Evaluation}

\begin{figure*}
    \centerline{\includegraphics[width=1.\textwidth]{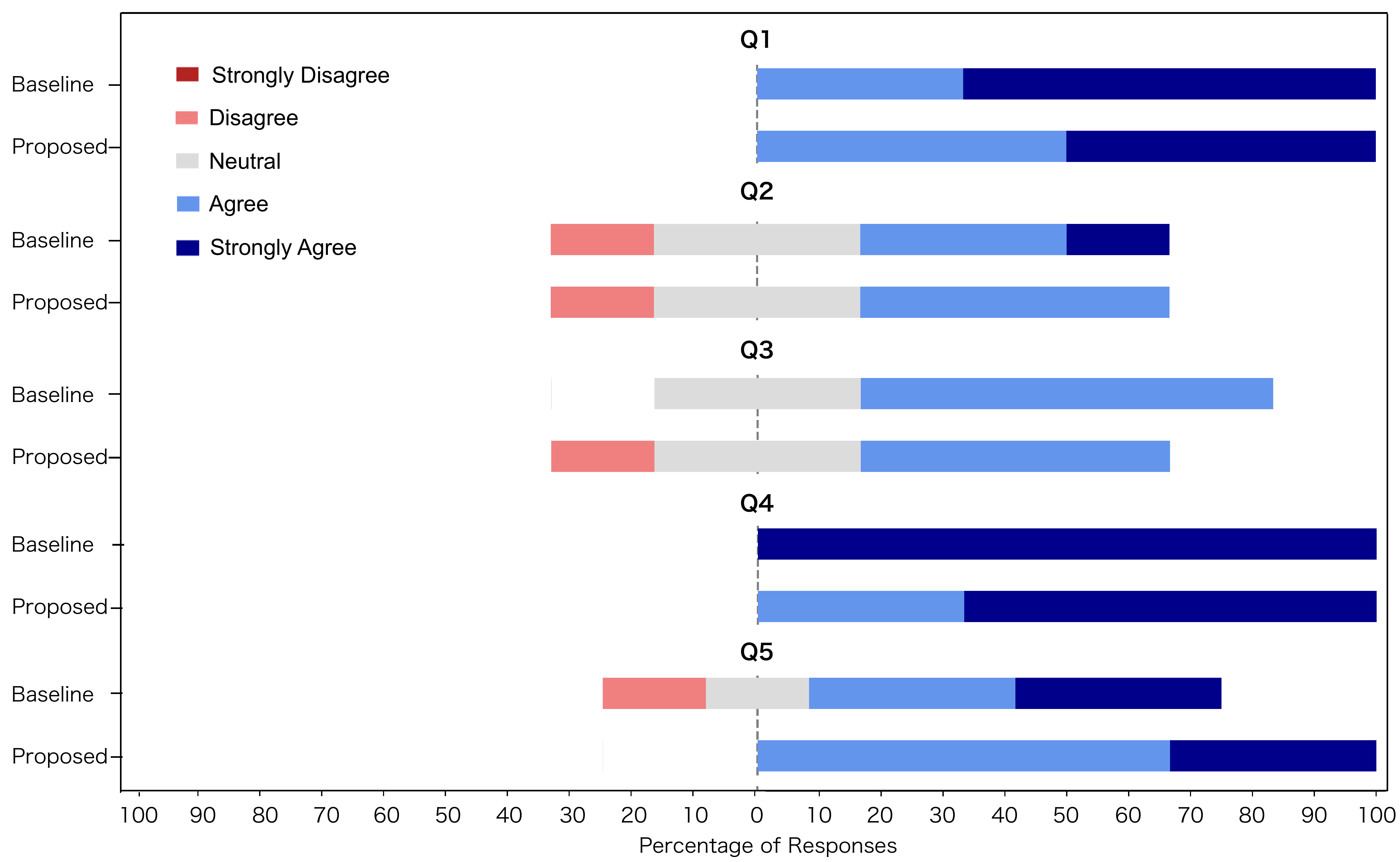}}
    \caption{Likert scale evaluation items (Q1 to Q5) for comparison between the baseline system and \systemname (proposed).}
    \label{fig:q1_q5_imla}
\end{figure*}

\begin{figure*}
    \centerline{\includegraphics[width=1.\textwidth]{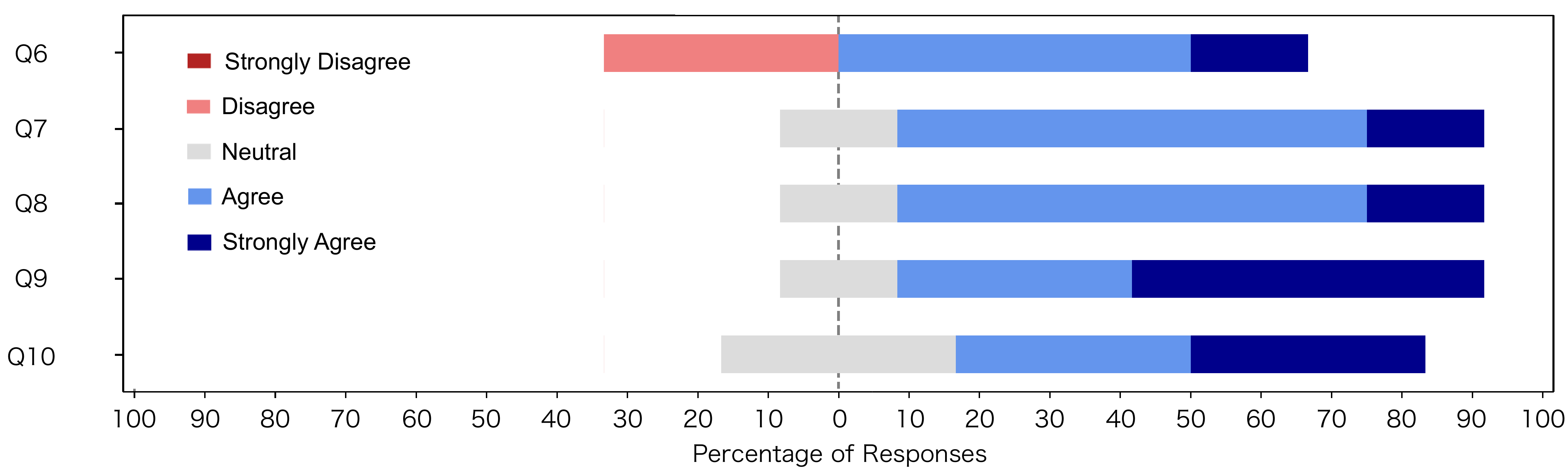}}
    \caption{Likert scale evaluation items specific to \systemname (Q6 to Q10).}
    \label{fig:q6_q10_imla}
\end{figure*}

Fig.~\ref{fig:q1_q5_imla} shows the results of Q1 to Q5, which are the Likert scale questionnaire items comparing \systemname and the baseline. 
Although \systemname includes additional interaction with the agents compared to the baseline, the responses to Q1, Q2, Q3, and Q4 suggest that this added complexity did not negatively impact usability. 
Statistical analysis (Wilcoxon signed-rank test) found no significant differences between the systems for any of these items. 
Interestingly, while participants gave similar self-assessments of their task performance success regardless of the system used, the third-party evaluation results (presented later) show \systemname to be more beneficial for users. 
This discrepancy suggests that users may not consciously recognize \systemname's advantages within their own awareness.

% % 定性的な方。自由記述とアンケート
% % システムに関する肯定的、否定的な発言のまとめ
Fig.~\ref{fig:q6_q10_imla} shows the results for Q6 to Q10, which are items specific to \systemname, reflecting participants' impressions of the agents' advice. 
Participants generally reacted positively to all items, indicating the meaningful role of the agents. 
This positive reception was further elaborated in participants' free-form responses. 
For instance, P4 appreciated the agents' clear guidance: ``\textit{It clearly explained which images to use for testing and what types of images to add (to the training data) to improve the model, which was easy to understand.}''
P10 valued the collaborative nature of the assistance: ``\textit{I liked that it gave hints to make the questioner think, rather than providing direct answers to questions.}''
However, some participants noted areas for improvement, such as P1's observation about repetitive advice: ``\textit{While the chat responses were often helpful, there were times when it kept repeating the same things. I wanted different advice as well.}''
% P1: チャットの回答が役にたつことが多々あったが、たまにさっきと同じことしか言わなくなる時があった。違うアドバイスも欲しかった。
% P4: どのような画像でテストを行えば良いかや、どのような画像を追加するとよりよいモデルになるかを明確に教えてくれてわかりやすかった。
% P10: 質問に対して全て答えを出すのでなく、ヒントを与えることで質問者に考えさせるのが良かった

% 具体的にどういうアドバイスをされたりどういう考えにさせられたか（主にインタビューからの引用で）
The follow-up interviews revealed diverse ways in which participants leveraged the MLLM-based agents to develop their classification strategies. 
Participants demonstrated three main patterns of agent interaction: hierarchical refinement of categories, knowledge extraction for better understanding, and adoption of abstract categorization principles. 
For example, P5 illustrated the hierarchical approach: ``\textit{For animal classification, since there are many images in the broad category of mammals, I asked the chat how to divide them further. 
Eventually, I categorized them into land, arboreal, and aquatic animals while regularly checking with the agent if this categorization was appropriate.}''
In terms of knowledge extraction, P10 effectively used the agent as an information source: ``\textit{The agent was helpful when I lacked prior knowledge. 
For animal classification, I could use it like Wikipedia, asking questions such as whether snakes are mammals or reptiles.}''
Regarding abstract categorization, P7 benefited from the active agent's suggestions: ``\textit{Looking at the advice, I realized that sometimes creating two categories like `A' and `not A' is more effective than dividing things into many detailed categories.}''
These diverse interaction patterns demonstrate how participants actively engaged with the agents to develop more sophisticated categorization strategies.

\subsubsection{Third Party Evaluation}
\begin{figure*}
    \centerline{\includegraphics[width=1.\textwidth]{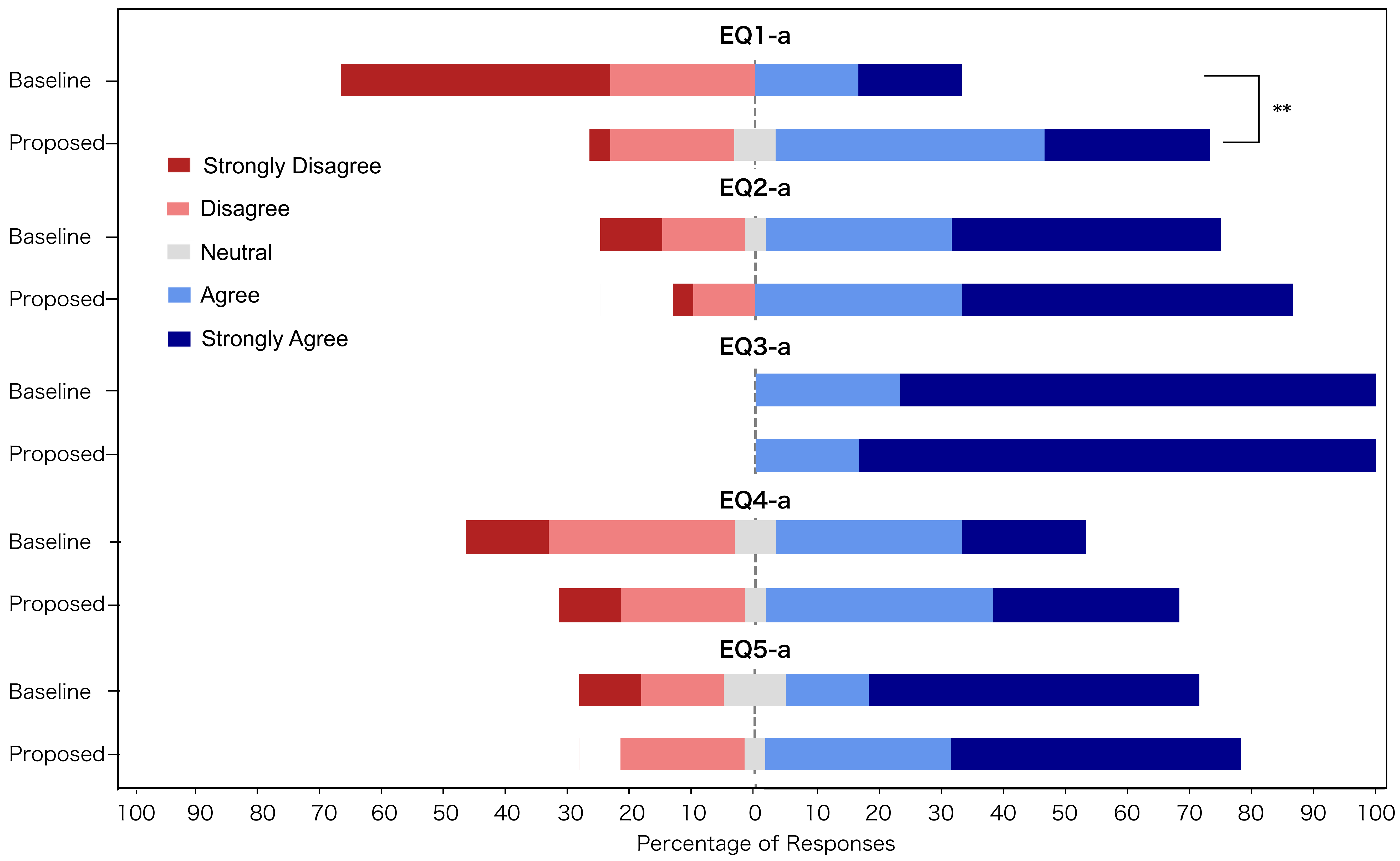}}
    \caption{Likert scale items in the third-party evaluation for comparison between the baseline system and \systemname (proposed).}
    \label{fig:third_party_imla}
\end{figure*}

The third-party evaluation was conducted by five practitioners (four males and one female, ages 19 to 38, M $= 29.00$, SD $= 7.18$) to assess the quality of participants' task formulation and category definition. 
The evaluation results, shown in Fig.~\ref{fig:third_party_imla}, consistently indicated higher performance for \systemname compared to the baseline system across multiple metrics.

% EQx-aについてこちらにまとめ
For the primary evaluation metrics (EQ1-a to EQ5-a), \systemname demonstrated consistently higher scores: task formulation accuracy (EQ1-a: $0.7$ vs. $-0.6$), category appropriateness (EQ2-a: $1.23$ vs. $0.83$), image-category matching (EQ3-a: $1.83$ vs. $1.77$), category completeness (EQ4-a: $0.57$ vs. $0.13$), and category distinction (EQ5-a: $1.03$ vs. $0.87$).
Statistical analysis using the Mann-Whitney U test, comparing the evaluations from each of the five evaluators separately, revealed a significant difference in task formulation accuracy (EQ1-a, $p = 0.002$). 
This finding is particularly noteworthy as it indicates that while participants' self-evaluations were similar between systems, \systemname users actually achieved more accurate task formulation according to objective analysis.

% EQx-bについてこちらにまとめ
The detailed metrics (EQ1-b to EQ5-b) further support \systemname's effectiveness in helping users define better categories.
Users of \systemname created more appropriate category names (EQ1-b: M $= 3.27$, SD $= 1.15$ vs. M $= 1.7$, SD $= 1.28$), better met task requirements (EQ2-b: M $= 4.1$, SD $= 1.58$ vs. M $= 3.27$, SD $= 2.00$), and achieved better image-category matching (EQ3-b: M $= 5.43$, SD $= 1.90$ vs. M $= 4.67$, SD $= 2.53$). 
While both systems showed similar performance in identifying missing categories (EQ4-b: M $= 1.4$, SD $= 0.82$ vs. M $= 1.38$, SD $= 0.47$), \systemname showed a slight improvement in reducing overlapping categories (EQ5-b: M $= 0.60$, SD $= 0.44$ vs. M $= 0.74$, SD $= 1.02$), indicating better category distinction.

\subsubsection{Interaction Log Analysis}
% 各機能を何回使ったかの簡単な統計
Analysis of participants' interaction logs revealed diverse patterns in how users engaged with the system's guidance features.
When using \systemname, participants actively sought assistance through multiple channels.
They directly messaged the passive agent more frequently during the directed task (M = $5.67$, SD = $3.08$) compared to the open-ended task (M = $3.83$, SD = $3.25$).
The active agent provided consistent guidance across both tasks, offering advice approximately 14 times per task (open-ended: M = $13.83$, SD = $3.76$; directed: M = $14.17$, SD = $3.82$).
Participants also utilized the ``Ask'' buttons differently between tasks, with more frequent inquiries about both categories and inference results during the open-ended task (categories: M = $4.17$, SD = $4.83$; inference: M = $7.00$, SD = $10.12$) compared to the directed task (categories: M = $1.33$, SD = $1.37$; inference: M = $3.50$, SD = $4.04$). 
These patterns suggest that participants adapted their interaction strategies based on task requirements.
% By examining the participants' interaction logs, we can infer how various types of guidance influenced them from the agent. 
% The six participants using \systemname sent messages to the passive agent an average of $3.83$ times (SD $= 3.25$) for the open-ended task and $5.67$ times (SD $= 3.08$) for the directed task.
% They also received advice from the active agent $13.83$ times (SD $= 3.76$) for the open-ended task and $14.17$ times (SD $= 3.82$) for the directed task.
% Regarding the ``Ask'' button, participants inquired about individual categories $4.17$ times (SD $= 4.83$) for the open-ended task and $1.33$ times (SD $= 1.37$) for the directed task.
% They asked about inference results $7.00$ times (SD $= 10.12$) for the open-ended task and $3.50$ times (SD $= 4.04$) for the directed task.
% These statistics indicate that participants utilized various system features evenly and interacted with the agents differently.

% Proposed systemを使った6人の被験者の対話ログを定性的に解析して、どういう対話をしてどういうアドバイスを受け取った結果どういう教師データになった...というのを時系列でつらつらと書いていく
% 動物タスクも、Caltech-101タスクも。両方について6人がどういう時系列でMLLMとやり取りしたかのサマリーを描く

Detailed analysis of user interactions revealed several distinct patterns in how participants refined their task formulations through agent collaboration. 
One common pattern involved the evolution from broad to specific categorizations.
For example, in the open-ended task, P10's initial general categories like ``insects'' and ``birds'' evolved into more specific classifications like ``butterflies'' and ``sparrows'' following agent guidance. 
Similarly, in the directed task, P7's broad concepts of ``money'' and ``vehicles'' expanded to include more nuanced categories like ``electronic devices'' and ``clothing'' after the agent identified potential classification gaps.
% Observing the interactions between participants and the assistant revealed several patterns in how the assistant supported task formulation. 
% In the open-ended task, the assistant's support manifested in three main ways. 
% First, it helped users refine overly broad categorizations: when P10 proposed general categories like ``insects'' and ``birds,'' the assistant suggested subdividing them into specific species like ``butterflies'' and ``sparrows.''
% Second, it offered alternative classification approaches based on user preferences: when P11 considered taxonomic classifications like ``mammals,'' the assistant proposed behavioral categories like ``carnivores, herbivores, omnivores.''
% Third, it provided strategic guidance on model evaluation, as seen when encouraging P14 to test with unexpected animal images to assess model limitations.

Beyond simple refinement, many participants demonstrated conceptual evolution through iterative dialogue with the agents. 
For instance, P11's shift from traditional taxonomic classifications to behavior-based categories (from ``mammals'' to ``carnivores, herbivores, omnivores'') exemplifies how agent interactions could inspire alternative classification approaches. 
This evolution was particularly evident in the directed task, where participants like P16 progressed from simple binary classifications (``hand-used items, foot-used items'') to more sophisticated systems after considering the agent's feedback about edge cases.
P18's development from a basic ``wearable/non-wearable'' distinction to a comprehensive system including ``vehicles, consumables, flora, and fauna'' further illustrates this pattern of conceptual growth.
% For the directed task involving object categorization, the assistant demonstrated flexibility in supporting different user approaches while ensuring alignment with task requirements. 
% Some users, like P7 and P10, started with broad categories and refined them through dialogue 
% For instance, P7's initial suggestions of ``money'' and ``vehicles'' were complemented by the assistant's additions of ``electronic devices'' and ``clothing.'' 
% Others, like P16, began with specific binary classifications (``hand-used items, foot-used items'') and evolved the approach based on the assistant's suggestions to consider edge cases. 
% This evolution was particularly evident in P18's case, where the assistant guided her from a simple binary classification (``wearable/non-wearable'') to a more comprehensive system including ``vehicles, consumables, flora, and fauna.''

These interaction patterns demonstrate the agents' effectiveness in providing multi-layered support through complementary strategies: direct category suggestions, refinement guidance, and performance evaluation advice. 
The agent's ability to adapt their support style—from offering specific categorical suggestions to providing strategic advice about edge cases—enabled participants to develop more robust classification systems while maintaining their autonomy in the decision-making process.
This balance between guidance and user agency appears to be a key factor in helping users develop more appropriate training data, as evidenced by the third-party evaluation results.

\subsubsection{The Effect of Multimodality of \systemname}
% 結果はwikiにメモっていく（GitLabの10月分 or 11月分のwikiページ参照）

\begin{figure*}
    \centerline{\includegraphics[width=\textwidth]{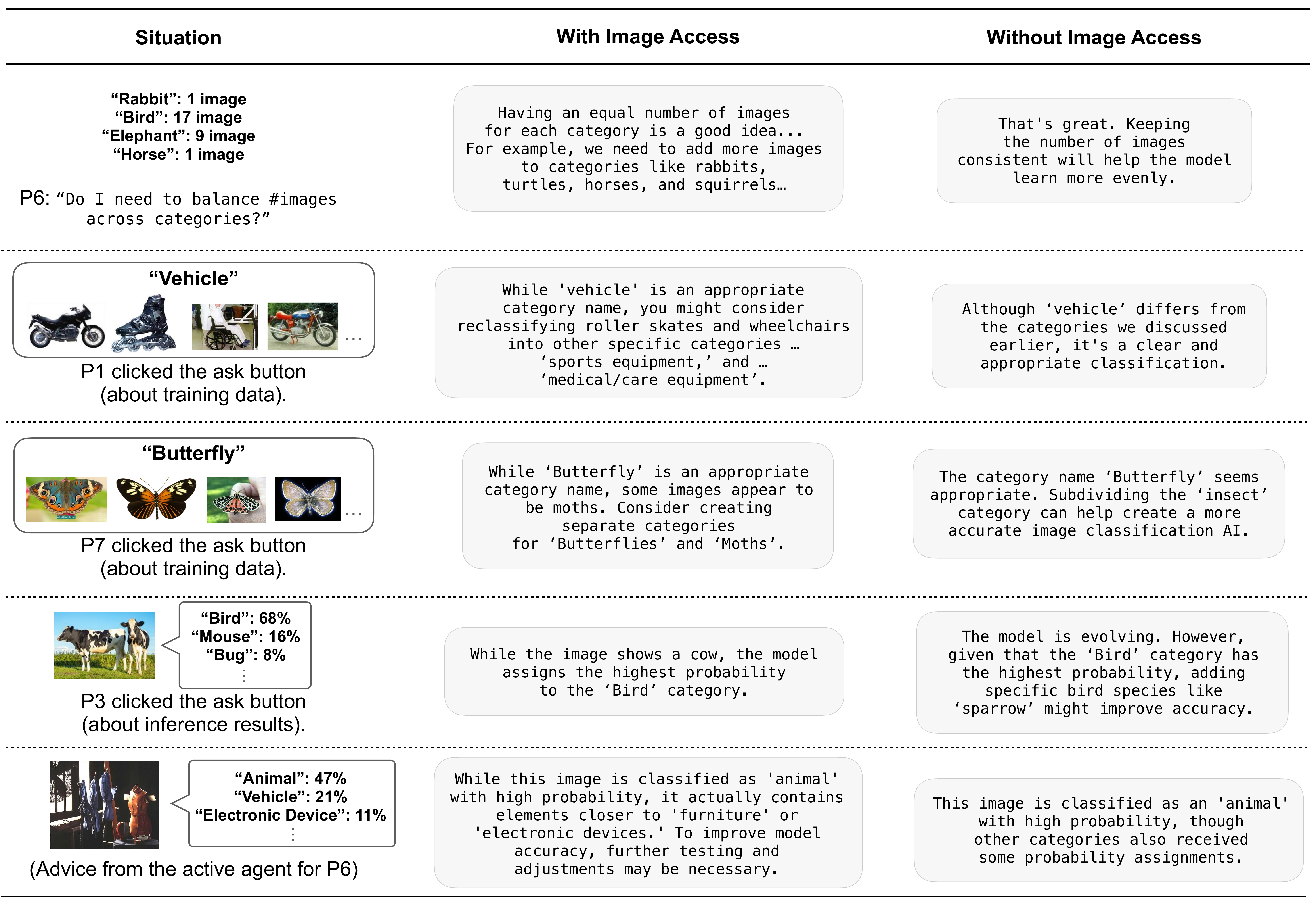}}
    \caption{To explore the potential benefits of MLLM's image recognition capabilities, we compared advice generated with image access to responses from text-only LLMs in identical scenarios. We analyzed both actual advice provided by the MLLM (With Image Access) and hypothetical responses from a text-only LLM (Without Image Access).}
    \label{fig:effect_multimodality}
\end{figure*}

%マルチモーダルじゃなかった世界線の場合どういう返答になるかを、promptログを元に復元してみる
%それを実際のログと比較してみる→MLLMであることのメリットを明らかにする
In \systemname, adopting MLLMs enables the system to provide more sophisticated advice through image understanding capabilities.
While text-only LLMs could assist users in basic task formulation, the ability to reference images significantly enhanced the quality and specificity of guidance.
To demonstrate these enhancements, we analyzed how the agents' outputs differ between cases with and without image access.

We show some examples in Fig.~\ref{fig:effect_multimodality}. 
The most notable difference was the agents' ability to provide specific guidance in category definition and refinement. 
When P6 asked about balancing image numbers across categories, the MLLM could offer precise guidance: ``\ctext{\texttt{Having an equal number of images for each category is a good idea... For example, we need to add more images to categories like rabbits, turtles, horses, and squirrels.}}'''
Similarly, when P1 created a ``vehicle'' category, the image-aware agent provided detailed feedback for refinement: ``\ctext{\texttt{While `vehicle' is an appropriate category name, you might consider reclassifying roller skates and wheelchairs into other specific categories.
Roller skates could go under `sports equipment,' and wheelchairs under `medical/care equipment.' The other items, like cars, motorcycles, airplanes, helicopters, and boats, fit well in the `vehicle' category.}}'''
In contrast, without image access, responses remained generic and non-specific: ``\ctext{\texttt{Keeping the number of images consistent will help the model learn more evenly}}''' and ``\ctext{\texttt{Although `vehicle' differs from the categories we discussed earlier, it's a clear and appropriate classification.}}'''

% （20241126）下の2段落を、上のように一段落分にまとめ
% We show some examples in Fig.~\ref{fig:effect_multimodality}.
% The most notable difference was the agents' ability to provide specific content references. 
% When P6 asked about balancing image numbers across categories, the MLLM could offer precise guidance: ``\ctext{\texttt{Having an equal number of images for each category is a good idea... For example, we need to add more images to categories like rabbits, turtles, horses, and squirrels.}}'''
% In contrast, without image access, responses remained generic: ``\ctext{\texttt{Keeping the number of images consistent will help the model learn more evenly.}}'''

% The MLLM also demonstrated superior capability in category refinement suggestions.
% When P1 created a ``vehicle'' category, the image-aware agent provided detailed feedback: ``\ctext{\texttt{While `vehicle' is an appropriate category name, you might consider reclassifying roller skates and wheelchairs into other specific categories. Roller skates could go under `sports equipment,' and wheelchairs under `medical/care equipment.' The other items, like cars, motorcycles, airplanes, helicopters, and boats, fit well in the `vehicle' category.}}'''
% Without image access, the response was non-specific: ``\ctext{\texttt{Although `vehicle' differs from the categories we discussed earlier, it's a clear and appropriate classification.}}'''

The MLLM's image understanding capabilities also enhanced error detection and classification analysis.
When P7 sought advice about their ``butterfly'' category, the MLLM could identify specific taxonomic issues: ``\ctext{\texttt{While `butterfly' is an appropriate category name, some images appear to be moths rather than butterflies. 
Consider creating separate categories for `butterflies' and `moths.'}}'''
Similarly, during model testing, when P3 tested their model with a cow image, the image-aware agent could precisely identify misclassification issues: ``\ctext{\texttt{While this image shows a cow, the model assigns the highest probability to the `bird' category.}}'''
In contrast, without image access, the agents could only provide superficial feedback such as ``\ctext{\texttt{The category name `butterfly' seems appropriate}}''' or irrelevant responses like ``\ctext{\texttt{These results show the model is evolving.}}'''

% Error detection and correction became more precise with image access. 
% For instance, when P7 sought advice about their ``butterfly'' category, the MLLM identified specific issues: ``\ctext{\texttt{While `butterfly' is an appropriate category name, some images appear to be moths rather than butterflies. Consider creating separate categories for `butterflies' and `moths.'}}''' 
% The text-only version could only offer superficial feedback: ``\ctext{\texttt{The category name `butterfly' seems appropriate.}}'''

% The MLLM provided more meaningful classification feedback during model testing. 
% When P3 tested their model with a cow image, the image-aware agent could identify misclassification issues: ``\ctext{\texttt{While this image shows a cow, the model assigns the highest probability to the `bird' category.}}'''
% It then suggested strategic improvements.
% Without image access, responses were often irrelevant: ``\ctext{\texttt{These results show the model is evolving.}}'''

Finally, the MLLM offered more detailed structural suggestions by analyzing inference results.
When P6's model classified an image, the active agent provided content-aware feedback: ``\ctext{\texttt{While this image is classified as `animal' with high probability, it actually contains elements closer to `furniture' or `electronic devices.' To improve model accuracy, further testing and adjustments may be necessary. Consider introducing unexpected categories like adding `lion' to the animal category or `fruits' to the food category for better validation.}}''' 
Without image access, the response was limited to surface-level observations: ``\ctext{\texttt{This image is classified as an `animal' with high probability, though other categories also received some probability assignments.}}'''
These examples demonstrate how multimodal capabilities significantly enhance the quality and usefulness of the agent's guidance across various aspects of the task.

\section{Discussion}

\subsection{Key Findings}

% （第三者評価から）\systemnameがユーザのタスク定式化を容易にすること
The third-party evaluation results demonstrate that human-MLLM collaboration through \systemname effectively supports users in formulating tasks and creating well-aligned training data. 
This collaboration proved particularly effective in EQ1-a, where \systemname significantly outperformed the baseline system in category name definition. 
While users can interact with the MLLM through various means, the MLLM currently communicates only through text. 
This text-based collaboration naturally showed the strongest impact on linguistic aspects like category naming. 
Expanding the collaboration channels beyond text, such as enabling MLLMs to generate and present images for suggesting training or evaluation data, could potentially enhance the partnership's effectiveness across other evaluation criteria.

% （ユーザビリティ定量評価から）\systemnameはMLLM機能を余計に追加しているが、ユーザビリティとしては落ちることがなく、むしろユーザはMLLMをappreciateしていること
While the system's effectiveness is clear from the evaluation results, the quantitative and qualitative usability results further reveal that users view the MLLM as a valuable collaboration partner rather than a burden.
Despite the additional burden of agent interactions compared to the baseline system, users consistently appreciated the collaborative experience. 
This positive response stems from the balanced nature of the human-MLLM partnership, where the MLLM acts as a peer in helping users concretize their ideas rather than providing directive instructions. 
Users might have resisted a top-down approach, but the MLLM's collaborative approach, maintaining a peer-like perspective and drawing out users' thoughts, proved effective.
The MLLM successfully enhanced both the user experience and task conceptualization, demonstrating the value of human-AI collaboration in ML.

% （インタビュー・定性から）\systemnameを通じて独自の発見にたどり着いたり、良い教師データとは何かを自分なりに考えさせるきっかけをユーザに与えられる、という副次的な効果があること
Beyond these immediate benefits in task completion and user experience, the human-MLLM collaboration in \systemname helps users develop deeper insights and refine their understanding of ML tasks.
When creating training data, users must consider various complexities, such as category overlaps (e.g., a wheelchair belonging to both ``handheld objects'' and ``vehicles'') and appropriate levels of abstraction for categories. 
Our observations show that participants were able to consider these aspects through collaborative dialogue with the MLLM thoughtfully. 
The interaction encourages users to pause and reflect, sometimes leading them to revise their initial assumptions or reconsider their approach to the ML model.
One of the most valuable aspects of this collaboration is how it guides users toward insights they might not have discovered independently.
This demonstrates how human-MLLM partnerships can enhance task completion and the entire problem-solving process.

% （画像モダリティ効果検証から）画像を参照した際はアドバイスの幅が広がり、定式化をアドバイスするだけじゃなくてML開発初心者へのヘルプとかチュートリアル的な雰囲気も帯びてくること。
Underpinning all these advantages is the fact that MLLMs' visual understanding capabilities significantly expand the range of collaborative possibilities beyond text-only interactions.
Through multimodal dialogue, users can freely explore various aspects of ML development, from basic concepts like ``\ctext{\texttt{What does model training and training data mean?}}''' to specific inquiries about model inference.
The MLLM's ability to analyze images helps users grasp fundamental ML concepts, including appropriate image quantity, variation, and category naming approaches.
However, realizing these benefits poses several challenges.
While this flexibility in human-MLLM interaction enables users to adapt \systemname to their individual learning needs, its effectiveness depends on thoughtful, prompt design and system architecture.
As a crucial consideration for further improvement, system designers must carefully consider how to structure these collaborative interactions to maximize their benefits while maintaining robustness.

\subsection{Limitation and Future Work}

% 画像モダリティを使い切れているとは言えない
The current implementation of \systemname extends human-MLLM collaboration beyond text to include image modality, enabling the MLLM to provide comprehensive guidance through both chat interactions and visual analysis. 
This multimodal collaboration allows the MLLM to offer specific suggestions for improving training data and inference images. 
The visual aspect of this partnership has significant potential for enhancement. 
Future iterations could incorporate techniques like style transfer~\cite{jing2019neural}, data augmentation~\cite{shorten2019survey}, or image generation~\cite{ramesh2021zero,isola2017image} to help MLLMs suggest new images while referencing existing ones. 
Enhanced image recognition capabilities could enable MLLMs to provide more detailed visual feedback, including specific areas needing improvement. 
The UI could evolve to support this collaboration by directly visualizing and annotating areas for improvement on-screen. 
While there are numerous possibilities for expanding this multimodal human-MLLM collaboration, implementing new features that align with \systemname's core objectives remains a future challenge.

% 本当はIMLシステムで求められてる手続きすらやらずに、対話システムで完結させたい
Beyond these technical limitations in multimodal interaction, a more fundamental challenge lies in the system's basic approach.
While \systemname enhances the conventional ML workflow through human-MLLM collaboration to help non-expert users formulate tasks correctly, research suggests that the ML workflow itself may present challenges for non-experts~\cite{kawabe2024technical}. 
An ideal system would enable ML without requiring specialized knowledge about training and inference, allowing users to train and evaluate models through natural dialogue with their MLLM partner.
However, achieving this vision would require expanding the MLLM's role beyond dialogue to execute specific operations, demanding a more sophisticated implementation of human-MLLM collaboration.

% ハイパラチューニング、モデルアーキテクチャ、データ拡張といった応用的な操作もサポートしたい
In addition to these architectural constraints, our current implementation is limited in its operational scope.
We focused on basic operations to provide a foundation for non-expert users, limiting the core workflow to creating training data, model training, and model evaluation. 
The human-MLLM collaboration was initially designed around these fundamental operations.
However, the practical development process involves more advanced operations such as hyperparameter tuning, model architecture modifications, and data augmentation.
Future frameworks should support collaborative work on these operations to serve non-expert users and AI/ML practitioners effectively. 
Extending human-MLLM collaborative ML to cover these complex aspects presents challenges, as it requires the MLLM to deeply understand model performance and user requirements.
This represents a crucial direction for advancing MLLM-supported ML systems.

% システム側が,MLLMの知識を使ってユーザにこんなデータをくれ、をサジェストできたらほんとは嬉しい
Finally, while \systemname helps users concretize their ideas through dialogues, prioritizing user intent from their initial vague needs. 
An alternative approach could involve the system independently defining desirable model states and requesting appropriate data from users to achieve these goals. 
Given the MLLM's common sense understanding, the system could potentially define characteristics of generally successful models.
This could enable an interactive machine teaching~\cite{ramos2020interactive} approach, where the system guides users toward predefined ideal outcomes. 
Such an implementation is feasible and represents another promising direction for the future development of human-MLLM collaborative ML.
These various limitations and future directions - from technical improvements in multimodal interaction to fundamental rethinking of the system's approach - highlight the rich potential for advancing human-MLLM collaborative ML systems while also emphasizing the complexity of creating truly effective tools for non-expert users.
\section{Conclusion}

In this paper, we introduced \systemname, a collaborative framework where users and MLLMs work together to formulate tasks appropriately and generate training data during ML model development. 
Through our implementation, the MLLM agents help users transform their vague requirements into concrete models while maintaining meaningful interaction. 
Our user study validated the effectiveness of this approach. 
It revealed that this human-MLLM partnership enabled users to formulate ML problems better, with users embracing the MLLM agents as valuable collaborators rather than viewing them as an additional burden.
Looking forward, as ML model training becomes increasingly important, frameworks that foster effective human-MLLM collaboration, like the one presented in this paper, will remain essential.
Future research should continue to explore ways to enhance these collaborative frameworks, focusing on helping users deeply consider their task requirements and create models that truly align with their specific challenges.

\section*{Acknowledgement}
This work was supported by JST SPRING, Grant Number JPMJSP2108.

\bibliographystyle{unsrt}
% \biblopgraphystype{junsrt}
\bibliography{main}

\appendix

\section{Prompts for the Passive Agent}

\begin{figure}
    \centerline{\includegraphics[width=.5\textwidth]{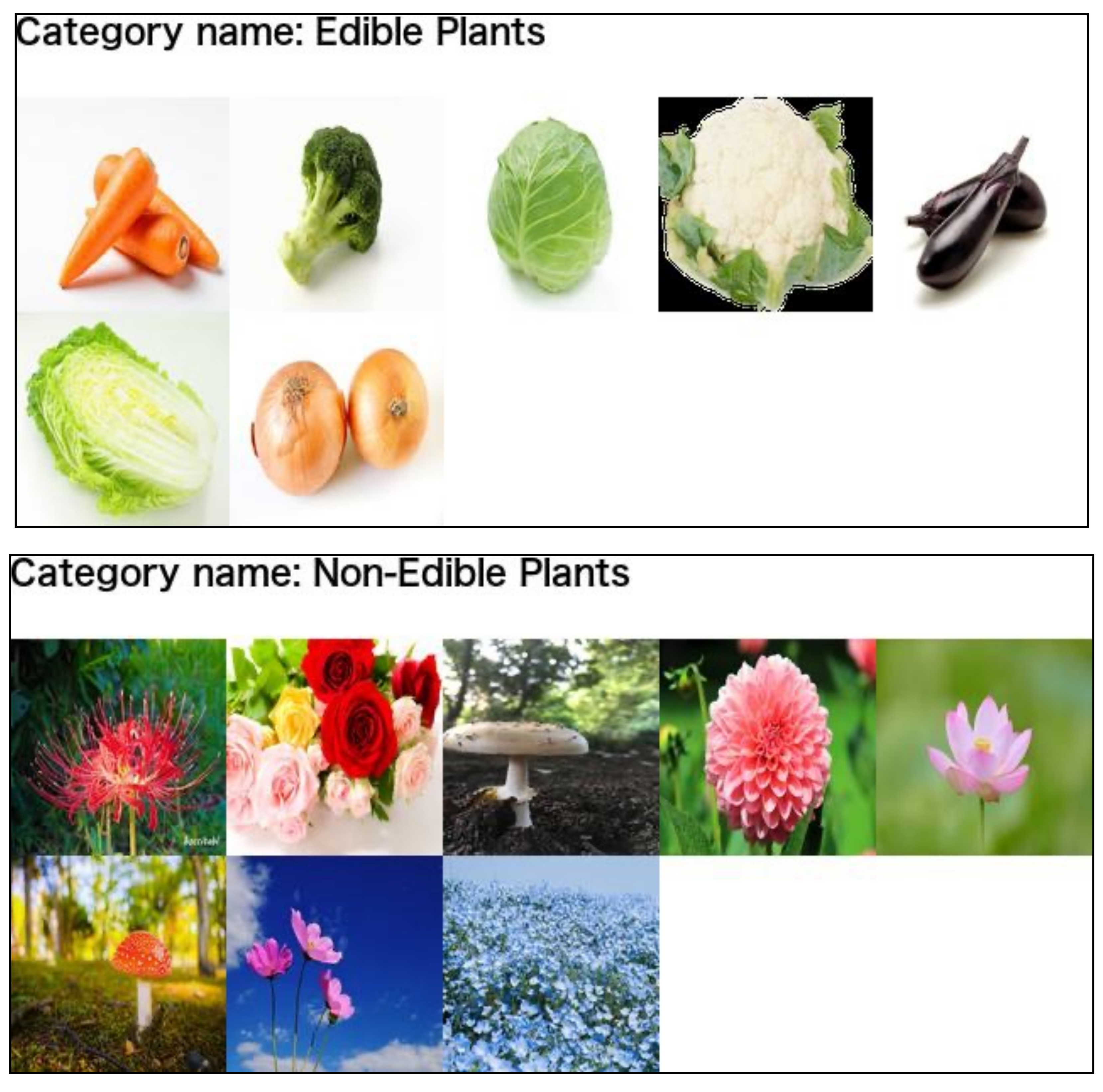}}
    \caption{Two examples of the image files sent to the server as user-defined training data. Each image file includes a category name and the associated images to the category name. Up to 50 associated images are randomly selected.}
    \label{fig:sent_images_imla}
\end{figure}

% This text is \ctext{\texttt{You can write something here, here here here here hexx this this this this this that he she it it it it.}}
% 2. If it's blockquote:
% \begin{blockquote}
    % \texttt{This is some text that will appear is a slightly smaller box with a gray background. You can use text inside the blockquote to change its font.}
% \end{blockquote}

We introduce the prompts sent to the passive agent. 
When user-created training data is included, it is attached to the prompt as image files. 
As shown in Fig.~\ref{fig:sent_images_imla}, each category is represented by a single image file containing both the category name and associated images. 
The associated images for each category name are randomly arranged, and a maximum of 50 are selected from those uploaded.

\subsection{System prompt}

\begin{blockquote}
\texttt{
Assist a user in creating an image classification model using their own data. The system allows users to define categories, upload images, train the model, and validate its accuracy. The model refers to the image features. The number of categories is flexible, with a maximum of 10. Start by addressing the user's possibly vague initial goal. Proactively engage with the user through dialogue, offering specific examples and suggestions to guide their machine learning problem formulation. Be mindful of potential misalignments such as inappropriate category names, insufficient category numbers, or unsuitable training images. Use concise dialogue to probe and clarify the user's needs, guiding their approach to problem formulation and data preparation. Suggest category names. Also, suggest testing with adversarial or ambiguous images during validation to uncover any overlooked categories that are critical to achieving the user's goals. The user has a limited number of images, so avoid giving advice about the quantity of data. Keep responses brief and natural, about one line. Always speak in \textbf{\textsf{(user\_selected\_language)}}.
}
\end{blockquote}

\subsection{User prompt}

Each user prompt is created and sent to the server when the user does an operation, such as sending a chat message.

\subsubsection{Chat input (if users have not defined training data yet)}

\begin{blockquote}
    \texttt{
    Please respond to the user's question/comment. The user's question/comment: `\textbf{\textsf{(user\_input)}}'.
    }
\end{blockquote}

\subsubsection{Chat input (after users defined training data)}
\begin{blockquote}
    \texttt{
    The attached images represent the training data defined by the user, showing the category names and the associated images (up to 50 images are displayed, even if there are more than 50). 
    Based on this, please respond to the user's question/comment. User's question/comment: `\textbf{\textsf{(user\_input)}}'.
    }
\end{blockquote}

\subsubsection{``Ask the assistant'' button (for a defined category)} 
\begin{blockquote}
\texttt{
Please refer to one of the categories from the training data. The category name is \textbf{\textsf{(user\_defined\_category\_name)}}, and the attached image displays the associated uploaded images (up to 50 images are shown, even if there are more). 
Review the appropriateness of the category name and images, both individually and in the context of previous discussions with the user.
}
\end{blockquote}

\subsubsection{``Ask the assistant'' button (for an inference result)} 
\begin{blockquote}
\texttt{
The attached image serves to validate the trained classification model. 
Inference results are: \textbf{\textsf{(inference\_result)}}
These results show all the category names defined by the user in the training data and the probabilities assigned to each category by the model. 
Encourage the user to introduce unexpected categories, like a lion in a dog and cat classifier or a fruit in a vegetable classifier, not only to test the model's limits but to inspire refinement in their problem formulation. 
This helps the user discover new potential categories and realize the importance of precise category definitions in classification models.
}
\end{blockquote}

Note that ``\ctext{\textbf{\textsf{(inference\_result)}}}''' refers to a JSON that stores the inference results of the trained model, where category names are keys and probabilities are values. 
For example: $\{$`dog'$: 30\%, $`cat'$: 20\%, $`bird'$: 50\%\}$.

\section{Prompts for the Active Agent}

\subsection{User Prompt}
Each user prompt is created and sent to the server at 60-second intervals.

\subsubsection{When users have defined training data} 

\begin{blockquote}
\texttt{
Advise a user to create an image classification model using their own data. 
The user interacts with the AI assistant during the model creation process, and all dialogues are recorded as \textbf{\textsf{(chat\_log)}}. 
Current training data is shown in the attached images, with up to 50 images per category. 
Focus on guiding the user to refine their vague ideas into a well-defined classification problem. 
Use concise dialogue to challenge and clarify the user's understanding of categories and model validation. 
Suggest category names. 
Also, suggest testing with adversarial or ambiguous images during validation to help identify any necessary but overlooked categories, refining the model's behavior to meet the user's specific goal. 
The user has a limited number of images, so avoid giving advice about the quantity of data. Keep communication clear, direct, and in \textbf{\textsf{(user\_selected\_language)}}.
}
\end{blockquote}

\subsubsection{When users have not defined training data yet} 
\begin{blockquote}
\texttt{
Advise a user to create an image classification model using their own data. The user interacts with the AI assistant during the model creation process, and all dialogues are recorded as \textbf{\textsf{(chat\_log)}}. 
Focus on guiding the user to refine their vague ideas into a well-defined classification problem. 
Use concise dialogue to challenge and clarify the user's understanding of categories and model validation. 
Suggest category names. 
Also, suggest testing with adversarial or ambiguous images during validation to help identify any necessary but overlooked categories, refining the model's behavior to meet the user's specific goal. 
The user has a limited number of images, so avoid giving advice about the quantity of data. Keep communication clear, direct, and in \textbf{\textsf{(user\_selected\_language)}}.
}
\end{blockquote}

\end{document}